\documentclass[useAMS,fleqn,usenatbib]{mnras}
\usepackage[T1]{fontenc}

\usepackage{graphicx,array}	
\usepackage{amsmath}	
\usepackage{amssymb}	
\usepackage{amsfonts}
\usepackage{array}
\usepackage{enumitem}
\usepackage{floatrow}
\usepackage[usenames,dvipsnames]{xcolor}
\usepackage{tabularx}
\usepackage[super]{nth} 
\usepackage{mathtools}
\usepackage{xspace} 
\usepackage[toc,page]{appendix}
\usepackage{afterpage}
\usepackage{subfig}
\usepackage{caption}
\usepackage{diagbox} 
\usepackage{makecell} 
\usepackage{bm} 
\usepackage{multirow}

\usepackage{layouts}

\usepackage{hyperref}

\def\aap{{ A\&A}}

\def\aj{{AJ}}

\def\apj{{ApJ}}
\def\apjl{{ApJL}}

\def\mnras{{MNRAS}}
\def\pasp{{PASP}}

\def\apjs{{ApJS}}

\def\CoRR{{CoRR}}

\newcolumntype{C}{>{\centering\arraybackslash}X}%

\newcommand{\unit}[1]{\ensuremath{\mathrm{\,#1}}\xspace}

\newcommand{\by}{\xspace\ensuremath{\times}\xspace}
\newcommand{\asec}{\unit{arcsec}}
\newcommand{\pix}{\unit{pix}}
\newcommand{\nm}{\unit{nm}}
\providecommand{\deg}{}
\renewcommand{\deg}{\unit{deg}}

\newcommand{\roughly}{\ensuremath{ {\sim}\,} } 
\renewcommand{\vec}{\textbfit} 
\newcommand{\eps}{\varepsilon} 
\newcommand{\Dpix}{{\Delta r}}
\newcommand{\D}[2][]{\frac{d #1}{d #2}} 
\newcommand{\RelFluxDiff}{\ensuremath{\Delta\Woline[0.85]{\mathcal{F}}^{\kern1pt ij}}}
\newcommand{\RelSNRDiff}{\ensuremath{\Delta\Woline[0.85]{\mathcal{S}}^{\kern1pt ij}}}

\makeatletter
\newsavebox\myboxA
\newsavebox\myboxB
\newlength\mylenA

\newcommand*\Woline[2][0.75]{%
    \sbox{\myboxA}{$\m@th#2$}%
    \setbox\myboxB\null
    \ht\myboxB=\ht\myboxA%
    \dp\myboxB=\dp\myboxA%
    \wd\myboxB=#1\wd\myboxA
    \sbox\myboxB{$\m@th\overline{\copy\myboxB}$}
    \setlength\mylenA{\the\wd\myboxA}
    \addtolength\mylenA{-\the\wd\myboxB}%
    \ifdim\wd\myboxB<\wd\myboxA%
       \rlap{\hskip 0.5\mylenA\usebox\myboxB}{\usebox\myboxA}%
    \else
        \hskip -0.5\mylenA\rlap{\usebox\myboxA}{\hskip 0.5\mylenA\usebox\myboxB}%
    \fi}
\makeatother

\DeclarePairedDelimiter\abs{\lvert}{\rvert}%
\DeclarePairedDelimiter\norm{\lVert}{\rVert}%

\makeatletter
\let\oldabs\abs
\def\abs{\@ifstar{\oldabs}{\oldabs*}}
\let\oldnorm\norm
\def\norm{\@ifstar{\oldnorm}{\oldnorm*}}
\makeatother

\title[Survey2Survey]{Survey2Survey: A deep learning generative model approach for cross-survey image mapping}
\author[Buncher, Sharma \& Carrasco Kind ]{Brandon Buncher$^1$\thanks{buncher2@illinois.edu},
Awshesh Nath Sharma$^{2}$,
and Matias Carrasco Kind$^{3,4,5}$\\
$^1$Department of Physics, University of Illinois, Champaign, IL 61820 USA\\
$^2$Department of Earth Sciences, Indian Institute of Technology Roorkee, Roorkee, Uttarakhand 247667 India\\
$^3$Department of Astronomy, University of Illinois, Urbana, IL 61801 USA\\
$^4$National Center for Supercomputing Applications, Urbana, IL 61801 USA\\
$^5$Center for Astrophysical Surveys, Urbana, IL 61801 USA}

\date{Accepted XXX. Received YYY; in original form ZZZ}

\pubyear{2020}

\begin{document}
\label{firstpage}
\pagerange{\pageref{firstpage}--\pageref{lastpage}}
\maketitle


\begin{abstract}
    During the last decade, there has been an explosive growth in survey data and deep learning techniques, both of which have enabled great advances for astronomy.  The amount of data from various surveys from multiple epochs with a wide range of wavelengths, albeit with varying brightness and quality, is overwhelming, and leveraging information from overlapping observations from different surveys has limitless potential in understanding galaxy formation and evolution.  Synthetic galaxy image generation using physical models has been an important tool for survey data analysis, while deep learning generative models show great promise.  In this paper, we present a novel approach for robustly expanding and improving survey data through cross survey feature translation.  We trained two types of neural networks to map images from the Sloan Digital Sky Survey (SDSS) to corresponding images from the Dark Energy Survey (DES).  This map was used to generate false DES representations of SDSS images, increasing the brightness and S/N while retaining important morphological information.  We substantiate the robustness of our method by generating DES representations of SDSS images from outside the overlapping region, showing that the brightness and quality are improved even when the source images are of lower quality than the training images.  Finally, we highlight several images in which the reconstruction process appears to have removed large artifacts from SDSS images.  While only an initial application, our method shows promise as a method for robustly expanding and improving the quality of optical survey data and provides a potential avenue for cross-band reconstruction.
\end{abstract}




\section{Introduction} \label{Intro}


The analysis of optical data at a wide frequency range collected by various astronomical surveys is a critical component used to study the origin and evolution of galaxies.  Data on galaxy shape \citep{GalShape} and luminosity \citep{GalLum1,GalLum2} in various bands provide information about the evolution of galaxies at different cosmic times.  As each band provides information about different characteristics of each object, stronger conclusions may be drawn from studies that incorporate data from a wide range of wavelengths.  While a large range of optical wavelengths is covered by most modern surveys, such as the Dark Energy Survey \citep[DES;][]{DESDR1} and the Sloan Digital Sky Survey \citep[SDSS;][]{SDSSDR7,SDSSCoadd}, the depth, the footprint, and signal-to-noise ratio (S/N) varies from survey to survey.  Inparticular, these will be vastly improved with future surveys like the  Legacy Survey of Space and Time (LSST) \citep{LSST}.  As a result, feature extraction in a particular band may be difficult in certain regions due to incomplete field coverage by surveys with high-quality data within that band.

\subsection*{Prior Work}

In order to understand the underlying galaxy formation model and physics behind galaxy properties, simulations are required to mimic observations; however, their systematics are computationally expensive.  Synthetic image generation of individual objects via deep learning is an alternative method for synthetic sky catalog generation that avoids the time and computational expense of other physically driven simulations.  Various neural network architectures have been used for this purpose, including variational autoencoders \citep{VAEGen1,VAEGen2,GalSimHub,AstroVaDEr} and generative adversarial networks (GANs) \citep{GANGen,LSSML}.  While these methods efficiently generate mock galaxy images, the accuracy of the output images depends on that of the input images.  As a result, their physical information is fundamentally limited by the quality of the survey data they are trained with.

In addition to image generation, autoencoders of various types have been used for a number of purposes in astronomy, including anomaly detection \citep{AEAnom} and object classification \citep{RGZoo,AstroVaDEr,SuperRAENN}.  GANs have also been utilized for feature extraction \citep{GANWL} and anomaly detection \citep{MLAnom}.

Two autoencoder architectures that are of particular importance for this work are convolutional autoencoders (CAEs) \citep{CAE} and denoising autoencoders (DAEs) \citep{DenoisingAE}.  CAEs have been utilized in astronomy for purposes including classification/feature extraction \citep{CAEMerge,CAELens,CAEHSI} and anomaly detection \citep{MLAnom}.  DAEs take as input an artificially corrupted input and are trained to reconstruct a distortion-free representation of that input.  DAEs are primarily used to eliminate noise from images \citep{DAEImg} and data \citep{DAEGravWave}, as well as for feature extraction \citep{DAEFX2,DAEFX1}.

One little explored alternative for improving the size and quality of survey datasets is through the use of feature transfer techniques across survey data.  A feature transfer model is trained to recognize differences between features in corresponding image pairs $\mathcal{X}$ and $\mathcal{Y}$ from datasets $\mathbb{X}$ and $\mathbb{Y}$.  Using an image from $\mathcal{X}'\in\mathbb{X}$ as input, the trained neural network can then be used to construct a representation of this image with the features characteristic of images in $\mathbb{Y}$.

In the context of astronomy and astrophysics, feature transfer learning using conditional GANs has recently found application for data analysis and feature extraction.  \citet{LineIntensity} developed a method to extract/reconstruct H$\alpha$ line intensity maps from noisy hydrodynamic simulation data.  In addition, \citet{GANWL} used feature transfer techniques to extract information from weak lensing maps.  However, other modified GAN architectures can be used for feature transfer learning; in particular, cycle-consistent generative adversarial networks (CycleGANs; these are described in Section \ref{Methodology}) are particularly suited for image analysis and generation.  Developed by \citet{CycleGAN} and \citet{Pix2Pix}, CycleGANs have been used for image-to-image translation (feature transfer between paired or unpaired sets of images) \citep{CycleSemantics,CycleUltra,CycleVid,CycleFor,MolGAN}.  However, there has been minimal exploration of generative models using feature transfer learning in astronomy and astrophysics.

Recently, \citet{AstroCycle} used image-to-image translation to reconstruct high-frequency noise patterns characteristic to different astronomical surveys.  The authors used several modified CycleGAN architectures with a semi-supervised training scheme using unpaired images to separate the signal and noise in images from two distinct surveys.  A noise emulator is then used to reconstruct the noise patterns from each survey.  The noise emulator can then be used to reconstruct images from a target dataset with said characteristic noise patterns.  While several of their models were successful at emulating noise patterns, training using unpaired images hindered the reconstruction of small-scale features of the signal.  Save for this work, the authors have been unable to identify any other use of CycleGANs in astronomy and astrophysics.

Methods other than feature transfer ones can hypothetically be used to generate representations of galaxies with altered parameters.  In particular, fader networks \citep{FaderNetwork,InvGAN} have been used by \citet{FaderGen} for the purpose of testing hypotheses about mechanisms that drive galaxy formation.  While this could be used as a method to transfer individual physical parameters of galaxies from one dataset to another, image reconstruction would not be feasible using this method because a large number of parameters must be known \textit{ab initio} to generate faithful representations of images in the target dataset.

We propose a novel method of feature transfer between galaxy surveys using CAEs and CycleGANs that can be used to expand galaxy image catalogs and can be adopted to multiple wavelengths and resolutions.  By training these architectures with images from DES DR1~\citep{DESDR1} paired with corresponding images from SDSS DR16~\citep{SDR16}, we demonstrate that information from DES images may be transferred to SDSS images, improving their S/N, contrast, and brightness.  We show that the synthetic DES images reconstructed from SDSS images share the same characteristics as the true DES images, and that this consistency is retained when performing reconstructions using images from a separate set of lower quality SDSS images which do not have a counterpart in the DES catalog.

While other works have demonstrated that variational autoencoders \citep{VAEGen1,VAEGen2,GalSimHub,AstroVaDEr} and GANs \citep{GANGen,LSSML} are effective at generating realistic synthetic galaxy images and improving the S/N, these models both train and validate using images from the same dataset.  Our method utilizes techniques that are generally similar to these; however, by using different data to train (SDSS) and validate (DES), we are able to generate false images that share the same morphological features of the SDSS images, but with a brightness and S/N more characteristic of the DES images.  Like other generative models, this can be used to increase the size of survey datasets; however, this method generates false representations of real observed galaxies.  This provides benefit when studying the properties of galaxies in a specific region that has not yet be covered by high quality surveys.  More importantly, transfer learning may allow for cross-band reconstruction: all surveys cover a limited range of wavelengths at sufficiently high quality for effective analysis, making feature extraction from particular bands impossible in certain regions of the sky.  By training using images with fewer bands than the validation data, a feature transfer-based generative model may be able to generate synthetic representations of galaxies with a greater range of wavelength bands than the input image.  This provides a method to allow more thorough analysis of galaxies in regions that lack sufficient band coverage.

In this work, we demonstrate the creation of \texttt{Survey2Survey}, a neural network architecture used to transfer features between SDSS and DES galaxy images that can be easily generalized to other optical surveys or even across multiple wavelengths.  The parameters of the SDSS and DES datasets used for training and validation are described in Section \ref{Data}.  In Section \ref{Methodology}, we detail the CAE and CycleGAN architectures used.  In Section \ref{Results}, we present qualitative and quantitative metrics of the accuracy of the reconstructed image, then summarize our findings in Section \ref{Conclusion}.

\section{Data} \label{Data}

In this section we describe the datasets used to carry out this study.  We focused on optical data from the SDSS and DES surveys and the overlapping region in the Stripe82 \citep{S82}.  All of the data used in this paper is publicly accessible via their respective websites.

All images consisted of three layers (one layer for each RGB channel), where the brightness of each pixel $P_i$ was represented by an 32-bit float, $0 \leq P_i \leq 1$.  Each SDSS image was 150\by150 \pix, and each DES image was 228\by228 \pix; the DES images were downscaled to match the dimensions of the SDSS images prior to training and reconstruction.  After the reconstruction and prior to the analysis, each three-layer 150\by150 \pix image was reduced to a single layer by averaging over the RGB channels.

\subsubsection*{SDSS} \label{SDSS}

SDSS images were captured by the Ritchey-Chr\`etien altitude-azimuth telescope \citep{RitChTel}, the Ir\`en\`ee du Pont Telescope \citep{IreneeTel}, and the NMSU 1-Meter Telescope \citep{NMSUTel}.  We selected a sample of galaxies in Stripe82 that overlapped with the DES footprint, and randomly sampled data from outside that region and within the northern cap for a total of 25,000 galaxies.  We chose galaxies with band Petrosian magnitude limits $14 < R < 17.77$, $z < 0.25$ and a resolution of $0.396 \asec/\!\pix$, using the galaxy flag produced by SDSS to select high confidence galaxy images.  Images of these galaxies were obtained from the SDSS cutout server\footnote{http://casjobs.sdss.org/ImgCutoutDR7}.

\subsubsection*{DES and Overlap Region} \label{DES}

DES uses the Dark Energy Camera \citep[DECam;][]{DECam} mounted at the Blanco 4m telescope at the Cerro Tololo Inter-American Observatory (CTIO) in Chile to observe $\roughly 5000 \deg^2$ of the southern sky in the $g$, $r$, $i$, $z$, and $Y$ broadband filters ranging from $\roughly 400\nm$ to $\roughly 1000\nm$ in wavelength.

We used images from the Dark Energy Survey DR1 release \citep{DESDR1}, which is comprised of over 10,000 co-added tiles of $0.534 \deg^2$ with a resolution of $0.263 \asec/\pix$ and a depth reaching S/N $\roughly 10$ for extended objects up to $i_{AB}\,\roughly 23.1$.

\begin{figure} 
    \centering
    \includegraphics[width = \columnwidth]{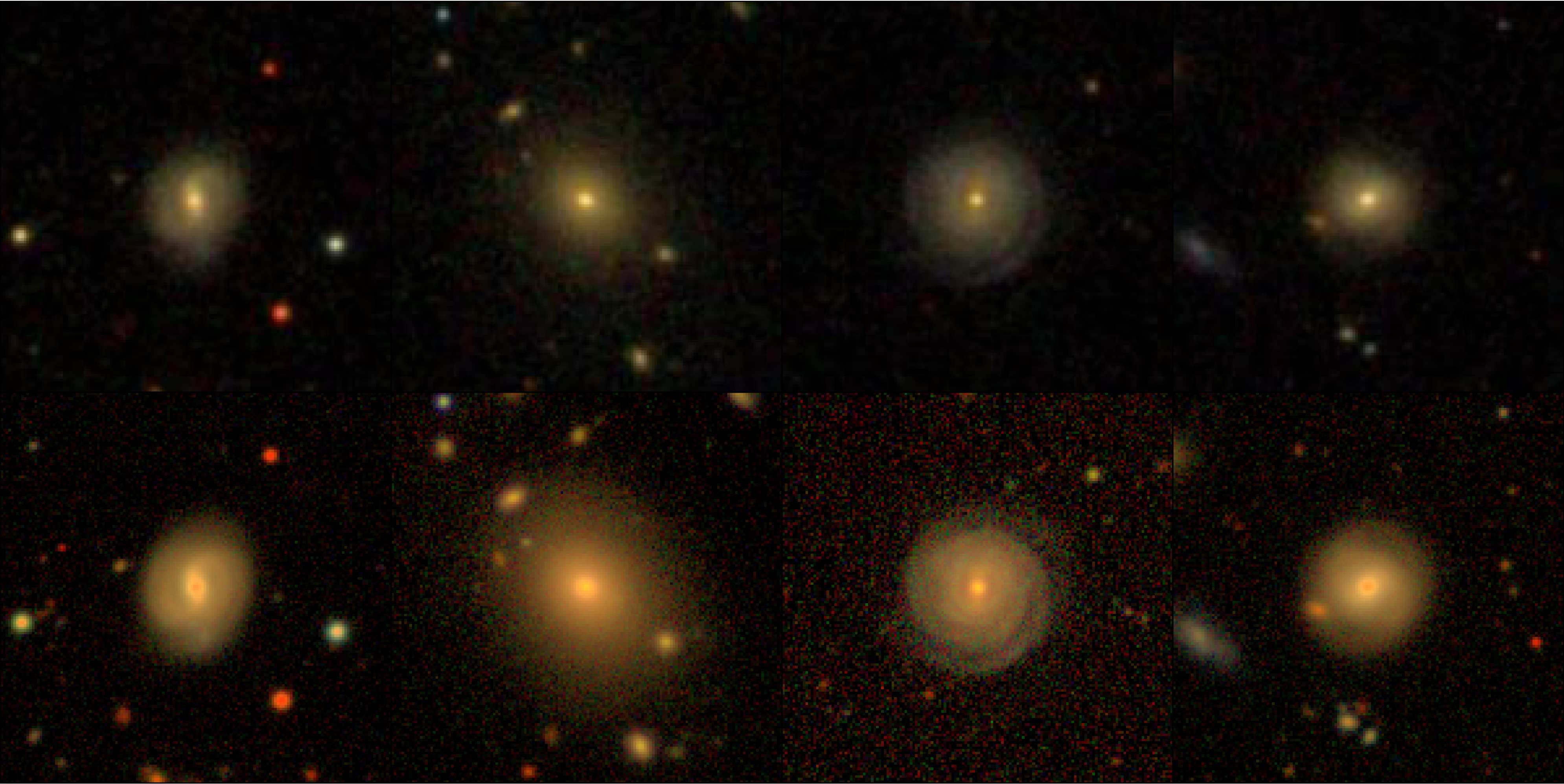}
    \caption{Sample images used from the Dark Energy Survey (DES) (top row) and Sloan Digital Sky Survey (SDSS) (bottom row) datasets.  More examples can be seen throughout the text.}
    \label{fig:DSImages}
\end{figure}

We selected DES galaxies using a combination of filtered criteria in terms of the concentration and error in the magnitude model as recommended\footnote{https://des.ncsa.illinois.edu/releases/dr1/dr1-faq} with $g < 17$ located in the Stripe 82 region \citep{S82} corresponding to roughly $300 \deg^2$ near the celestial equator.  We selected all images from Stripe 82 that have an SDSS counterpart \citep{SDSSDR7,SDSSCoadd}.  These images were obtained using the public DES cutout service\footnote{https://des.ncsa.illinois.edu/desaccess}.  We removed images with incomplete coverage and cleaned the images of anomalies and contaminants such as stars using visual inspection.  Each DES image was scaled to 150\by150 \pix to match the resolution of the SDSS images.  We aligned the orientation and central pixels of each DES/SDSS image pair, and the final RGB composite was generated using the \cite{RGBCCD} prescription in order to closely match the SDSS colors.  Figure \ref{fig:DSImages} shows examples of the galaxies selected, where we can see that the DES images appear brighter and more detailed than the SDSS images.

The overlap region was used for training and validation; each SDSS image in the overlap region had a DES counterpart.  In total, there were 5,538 RGB images in the overlap region.  5,000 SDSS/DES image pairs were used for training the models, while the remaining 538 were used as the validation dataset. Because of the large variation in the brightness and spatial extent of objects in the SDSS and DES datasets, we chose to use a training dataset that was $\approx 5\times$ larger than those used by both \citep{Pix2Pix} and \citep{CycleGAN}.

The external dataset, which consisted of 25,076 from outside of the Stripe 82 region, was used to provide evidence for the robustness of our methodology.  These images were fainter and of lower S/N than the training and validation datasets.

\section{Methodology} \label{Methodology}

Convolutional Autoencoders (CAE) \citep{CAE} and Cycle-Consistent Generative Adversarial Networks (CycleGAN) \citep{GAN} were used to generate synthetic galaxy images from the SDSS input images.  Since the images were scaled, rotated, and centered so that each pair of pixels in a given image pair corresponded with one another, minimizing the loss function used for both models corresponded with minimizing the pixel-to-pixel differences between the reconstructed image and the DES target image.  These two types of models differ in their implementation and objective function as described below.  We did not perform any methods that have traditionally been used to reduce overfitting and provide data augmentation, such as image rotations and translations, for either the CAE or CycleGAN.  Spatial transformations would have likely led to failure: as we intend to perform pixel-to-pixel translations, any misalignment of pixels would lead to the creation of an invalid mapping function.  While this may not cause an issue in many other cases, spatial transformations on SDSS and DES images could drastically reduce the accuracy of the mapping function given the small spatial extent of the signal region relative to the background in many images.  While this may have led to overfitting, the analysis of the external reconstructions provides evidence of the robustness of our method.  As an initial application of image-to-image translation for false image generation, we chose to minimize the number of factors that could affect the pixel-to-pixel map; future research should be dedicated to establishing methods to ensure that overfitting does not occur.

\subsection{Convolutional Autoencoders (CAE)} \label{CAE}

An autoencoder is a neural network architecture typically used for classification that is comprised of an encoder/decoder pair.  The encoder compresses data from an input image using one or more hidden layers to isolate important features from that image, generating a latent space representation of that image with lower dimensionality.  The decoder uses the information in the latent space to reconstruct a representation of the input image.  The autoencoder is trained to optimize a loss function to minimize the difference between the source and reconstructed image.  A convolutional autoencoder performs encoding and decoding using convolution filters: during the encoding stage, convolution filters are used to extract information from and decrease the dimensionality of the input image.  Additional convolution filters are used to map the latent space representation to a reconstruction of the input image.  Training is performed by iteratively modifying the weights of the convolution filters to minimize the differences between the source image and its reconstruction.

Our CAE was implemented using \texttt{Keras} \citep{Keras} using a \texttt{Tensorflow} \citep{TensorFlow} backend, and was run on a 32 GB Tesla V1000-PCIE GPU.  Training over the course of 100 epochs (a value chosen via early stopping) took $\roughly 30$ minutes.  Details about our architecture are shown in Table \ref{tab:CAEArch}.  We intentionally did not substantially decrease the dimensionality of the latent space of each layer because of the complexity of the images we aimed to reproduce.  The SDSS images were generally less bright and noisier, and objects from the DES dataset often had a greater number of pixels distinguishable from the background noise (i.e. the signal in DES images had a larger spatial extent) than the SDSS images, so it is unlikely that a low-dimensionality latent space would be capable of producing sufficiently detailed false images.

\begin{table}
    \centering
    \setlength\tabcolsep{4pt}
    \setlength\extrarowheight{2pt}
    \begin{tabular}{|c|c|c|c|}
        \hline
        \textbf{Stage} & \textbf{Output Shape} & \textbf{Activation} &  $\bm{N_{\textnormal{\textbf{P}}}}$ \\
        \hline
        \textbf{Input} & 150\by150\by3 & \multicolumn{2}{c|}{N/A} \\
        \hline
        \multirow{3}{*}{\textbf{Encoder}} & 150\by150\by128 & ReLU & 3584 \\
        {} & 150\by150\by64 & ReLU & 73792 \\
        {} & 150\by150\by32 & ReLU & 18464 \\
        \hline
        \multirow{4}{*}{\textbf{Decoder}} & 150\by150\by32 & ReLU & 9248 \\
        {} & 150\by150\by64 & ReLU & 18496  \\
        {} & 150\by150\by128 & ReLU & 73856 \\
        {} & 150\by150\by3 & Sigmoid & 3459 \\
        \hline
    \end{tabular}
    \caption{CAE architecture used for image reconstruction.  The initial input and final output images were 150\by150 \pix with 3 color channels; the output shape of each image in the encoder and decoder stages is length\by~width\by~no. filters.  Each row in the encoder and decoder stages represents a single convolution layer with the specified activation function; convolution was performed using 3\by3 kernel with a stride of 1 and zero padding.  The image passed to the subsequent convolution layer had dimensions corresponding to that row's output shape.  $N_{\textnormal{P}}$ is the number of training features in that layer.  The number of filters and activation functions used were chosen through manual tuning.}
    \label{tab:CAEArch}
\end{table}

The RGB data from each image was separated into three layers, each of which were used to generate a unique set of filters.  The encoder and decoder both consisted of three hidden layers, each of which filtered the image data from the previous layer using 150 3\by3 \pix convolution filters.  These filters were initialized using randomly generated weights.  Rectified Linear Unit (ReLU) activation functions were used for each layer of the encoder and decoder, and a sigmoid activation function was used during the final reconstruction phase.  For each epoch, the input image $\vec{x}_0$ was an image from the SDSS catalog, while the target image $\vec{x}_T$ was the same object taken from the DES catalog.  The difference between the reconstructed image $\vec{x}'_0$ and the target image was calculated using the mean squared error loss function

\begin{align}
    \mathcal{L}\left(\vec{x}'_0, \vec{x}_T\right) &= \norm{\vec{x}_T - \vec{x}'_0}
\end{align}

The \texttt{Adadelta} \citep{ADADELTA} optimizer was used to determine filter weights.  At the conclusion of 100 training epochs, the trained algorithm was used to reconstruct the DES validation images from their corresponding SDSS image.

\subsection{Cycle-Consistent Generative Adversarial Networks (CycleGAN)} \label{CycleGAN}

\begin{figure*} 
    \centering
    \begin{minipage}{\textwidth}
        \centering
        \includegraphics[width = \textwidth,keepaspectratio]{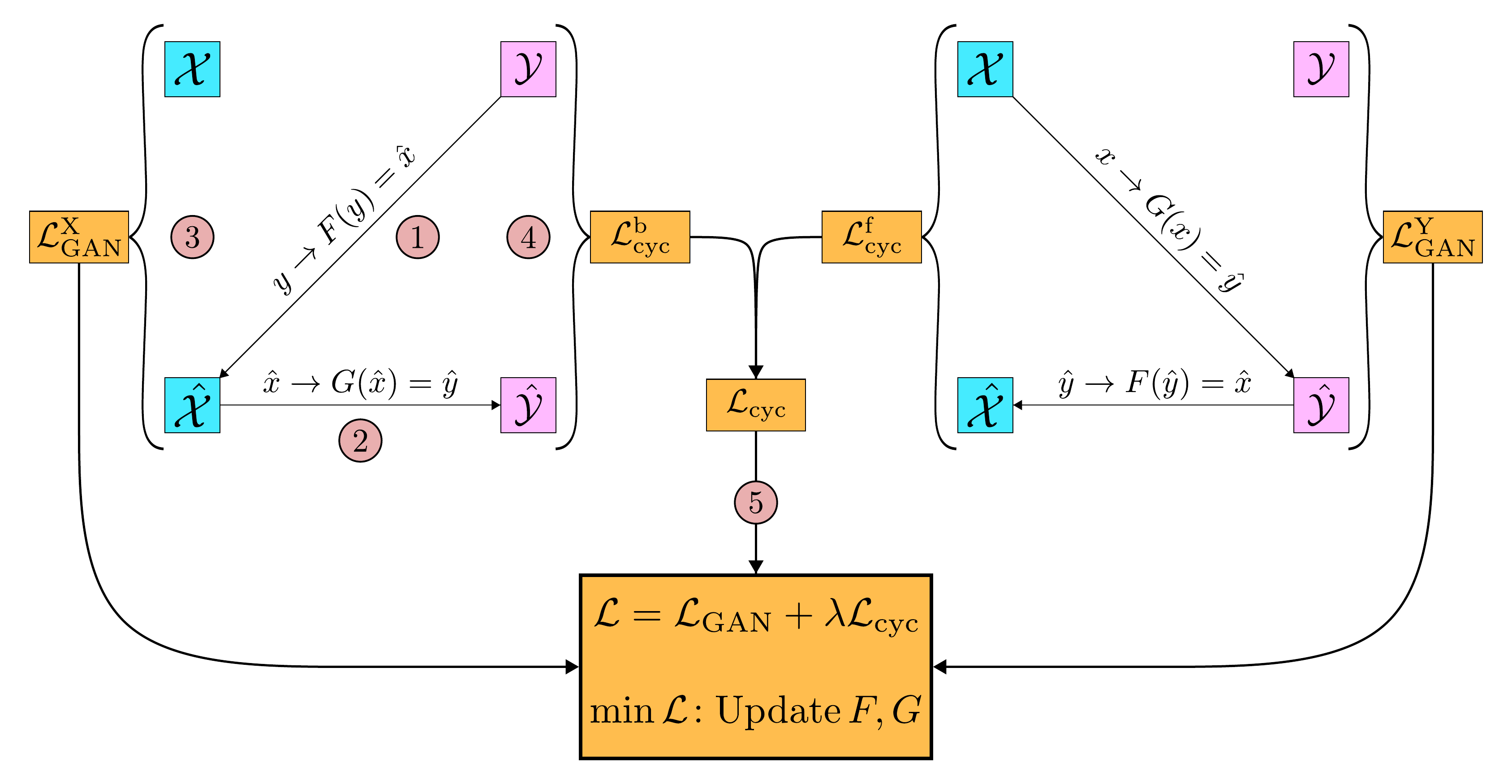}
    \end{minipage}
    \caption{A representation of the architecture of a CycleGAN.  \textbf{1.} A false image representation $\hat{x}$ is generated from $y$, a member of the target dataset, via the mapping function $F$.  \textbf{2.} $\hat{x}$ is mapped to a false image $\hat{y}$ via the mapping function $G$.  \textbf{3.} The GAN loss function $\mathcal{L}_{\textnormal{GAN}}^X$ for discriminator $D_{\mathcal{X}}$ is calculated by comparing $x$ and $\hat{x}$.  \textbf{4.} The backward cycle-consistency loss function $\mathcal{L}_{\textnormal{cyc}}^{\textnormal{b}}$ is calculated by comparing $\hat{y}$ to the true target image $y$ (the forward cycle-consistency loss function $\mathcal{L}_{\textnormal{cyc}}^{\textnormal{f}}$ is calculated similarly using $\hat{x}$ and $x$).  In our case, to calculate $\mathcal{L}_{\textnormal{cyc}}^{\textnormal{f}}$, we generate a DES representation of an SDSS image $x$, then use $F$ to generate a false SDSS representation of that image.  $\mathcal{L}_{\textnormal{cyc}}^{\textnormal{f}}$ quantifies the differences between the source SDSS image and the false SDSS image; its combination with $\mathcal{L}_{\textnormal{cyc}}^{\textnormal{b}}$ quantifies the error accumulated when the SDSS image completes a full ``cycle'' between the SDSS and DES image spaces ($\mathbb{X} \to \mathbb{Y} \to \mathbb{X}$).  \textbf{5.} These loss functions are combined with $\mathcal{L}_{\textnormal{GAN}}^Y$ and $\mathcal{L}_{\textnormal{cyc}}^{\textnormal{f}}$ to calculate the total loss function $\mathcal{L}$.  $F$ and $G$ are then updated to minimize $\mathcal{L}$.  This process is repeated to optimize the neural network.
    \label{fig:GANArch}}
\end{figure*}

A Generative Adversarial Network (GAN) \citep{GAN} is an unsupervised or semi-supervised generative model consisting of a generator $G$ and discriminator $D$.  $D$ is trained to distinguish between images from a training dataset of ``true'' images ($\mathcal{Y}$) and those generated by sampling from the latent space of $G$ ($\mathcal{X}$).  Backpropagation of error from $D$ is used to generate a map $g:\mathcal{X}\to\mathcal{Y}$ from the latent space of $G$ to the ``true'' image dataset by minimizing a loss function $\mathcal{L}(G, D, \mathcal{X}, \mathcal{Y})$.  After training, the GAN may be used to generate false images that replicate the features of $\mathcal{Y}$.

A CycleGAN \citep{CycleGAN,Pix2Pix} is a variation of a traditional GAN that minimizes cycle-consistency loss through the additional of a second generator/discriminator pair; a diagram of this architecture is shown in Fig. \ref{fig:GANArch}.  Images from $\mathcal{X}$ ($\mathcal{Y}$) are used to train discriminators $D_\mathcal{X}$ ($D_\mathcal{Y}$).  The generators $F:\mathcal{X}\to\mathcal{Y}$ and $G:\mathcal{Y}\to\mathcal{X}$ are trained to extremize the adversarial loss function $\mathcal{L}(H, D_\mathcal{Y}, \mathcal{X}, \mathcal{Y})$ for generator $G$, discriminator $D_\mathcal{Y}$, and datasets $\mathcal{X}$ and $\mathcal{Y}$.  For the purposes of this project, we chose to use the loss function used by \citet{CycleGAN}:

\begin{align} \label{eq:GLoss}
    \mathcal{L}_\textnormal{GAN}(G, D_\mathcal{Y}, \mathcal{X}, \mathcal{Y}) \kern2pt=\kern2pt &\mathbb{E}_{\kern0.5pt y\sim p_\textnormal{data}(y)}\left[\kern0.5pt\log D_\mathcal{Y}(y)\kern0.5pt\right] +\\ &\mathbb{E}_{\kern0.5pt x\sim p_\textnormal{data}(X)}
    \left[\kern0.5pt\log(1 - D_\mathcal{Y}(G(x))\kern0.5pt\right] \nonumber
\end{align}

for images $x\in\mathcal{X}$ and $y\in\mathcal{Y}$, where $p_\textnormal{data}$ is the true data distribution.  $G$ was trained to maximize $\mathcal{L}_\textnormal{GAN}$ ($\max_G\max_{D_\mathcal{Y}}\mathcal{L}_\textnormal{GAN}(G, D_\mathcal{Y}, \mathcal{X}, \mathcal{Y})$), while $F$ was trained to minimize it ($\min_F\max_{D_\mathcal{X}}\mathcal{L}_\textnormal{GAN}(F, D_\mathcal{X}, \mathcal{Y}, \mathcal{X})$).

To constrain the space of possible mapping functions, a CycleGAN optimizes $F$ and $G$ by minimizing the forward and backward cycle consistency error.  For images $x\in\mathcal{X}$ and $y\in\mathcal{Y}$, let $x' = F(G(x))$ and $y' = G(F(y))$.  Forward cycle consistency is achieved when the difference between $x'$ and $x$ is minimized (i.e. $F = G^{-1} + \eps_x$ for some small error $\eps_x$), indicating that the full translation cycle beginning in $\mathcal{X}$ reproduces a close approximation of $x$; backward cycle consistency is defined identically for images $y \in \mathcal{Y}$ and $G = F^{-1} + \eps_y$ for some small error $\eps_y$.  An optimized CycleGAN will simultaneously minimize the forward and backward cycle-consistency error; this is equivalent to ensuring that $F$ and $G$ are bijective inverses of one another, limiting the size of the set of possible mapping functions.  This improves the robustness of the neural network and decreases the amount of training required relative to many other GAN variations.

We note that for our particular case we used the architecture described in \cite{Pix2Pix} which is adapted from the unsupervised representation learning GAN architecture introduced in \cite{radford2016}. In particular, we highlight the use of a generator with skips and a Markovian discriminator.  These additions helped with the translation process and limited the GAN discriminator to high-frequency structures, reducing the potential for artifacts.

The cycle-consistency loss function $\mathcal{L}_\textnormal{cyc}(G, F)$ we used is defined as

\begin{align} \label{CLoss}
    \mathcal{L}_\textnormal{cyc}(G, F) =\kern3pt &\mathbb{E}_{x\sim p_\textnormal{data}(x)}\left[\kern0.5pt\abs{F(G(x)) - x}_1\right] + \\
    &\mathbb{E}_{y\sim p_\textnormal{data}(y)}\left[\kern0.5pt\abs{G(F(y)) - y}_1\right], \nonumber
\end{align}

where $\abs{A - B}_1 = \sum\limits_i\hspace{2pt}\abs{A_i - B_i}$ is the pixel-to-pixel $L^1$-norm between images $A$ (SDSS) and $B$ (DES).

The full loss function used for training $F$ and $G$ was

\begin{align}
    \mathcal{L}(G, F, D_\mathcal{X}, D_\mathcal{Y}) \kern2pt=\kern2pt &\mathcal{L}_\textnormal{GAN}(G, D_\mathcal{Y}, \mathcal{X}, \mathcal{Y}) \kern2pt+ \\ \nonumber
    &\mathcal{L}_\textnormal{GAN}(F, D_\mathcal{X}, \mathcal{Y}, \mathcal{X}) \kern2pt+ \\
    &\lambda\kern0.5pt\mathcal{L}_\textnormal{cyc}(G, F) \nonumber
\end{align}

for some parameter $\lambda$, which describes the relative importance of the optimization of the adversarial and cycle consistency errors.  For this work, we set $\lambda = 0.2$.

Image translation using a CycleGAN architecture provides benefit over a traditional GAN by constraining the allowed mapping functions by ensuring that the discriminator pair $F$ and $G$ are inverses.  This benefits the translation between noisy images by making sure that the differences in noise patterns in $\mathcal{X}$ and $\mathcal{Y}$ is taken into account, helping distinguish between the signal and noise more easily after training on many images.

\section{Results and Analysis} \label{Results}

Here, we demonstrate that we can transfer information from DES images to their SDSS counterparts, generating synthetic images that are brighter, of higher quality, and have less noise, yet retain the morphological information contained within the source image.  We begin with a qualitative analysis to understand properties of the reconstructed images, then quantify the brightness and noise level of the image datasets.  We then use correlations between the light profiles of the source and reconstructed objects to establish the small-scale differences between the datasets.  Finally, we combine this information with comparative quality assessments to establish that the image reconstruction process improves the image quality, brightens objects, and reduces background noise.  We provide evidence for the robustness of the reconstruction process by comparing the statistics of the validation and external datasets.

\subsection{Qualitative Analysis} \label{QualAnalysis}

\begin{figure*} 
    \centering
    \begin{minipage}{\textwidth}
        \centering
        \includegraphics[width = \textwidth]{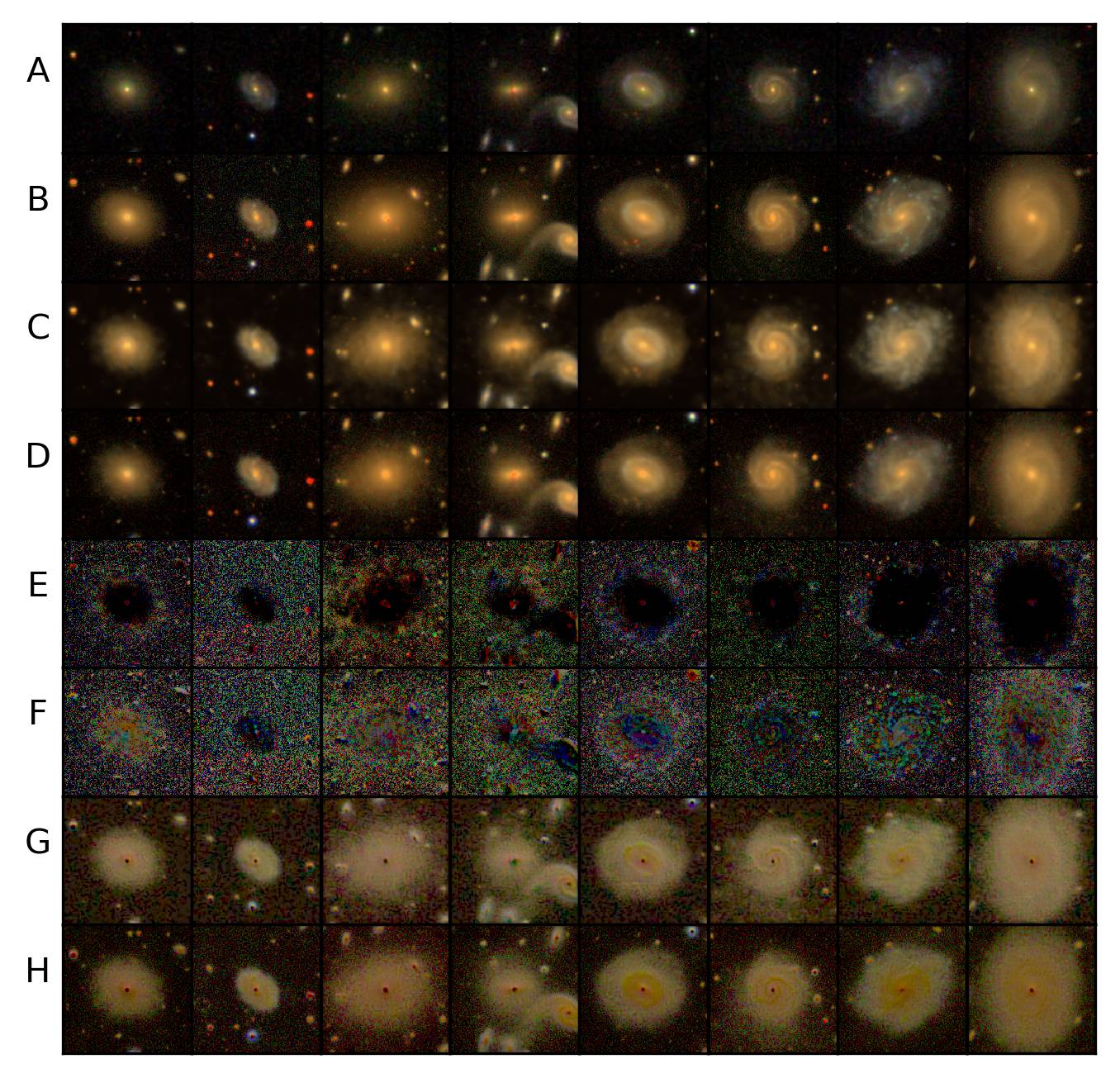}
        \end{minipage}
    \caption{Examples of galaxy images from the validation dataset (from the Stripe82 region).  Each column shows an SDSS galaxy (row A), its DES counterpart (row B), and the DES image reconstruction by the CAE (row C) and CycleGAN (row D) methods.  CAE and CycleGAN residuals (reconstruction - DES) are shown in rows E and F respectively, wile the CAE and CycleGAN pixel-to-pixel brightness increases (reconstruction - SDSS) are shown in rows G and H, respectively.  \textbf{Note that to increase visibility, images in rows E, F, G and H were artificially enhanced with a power law transform (\boldmath$P_i\,\to\,{P_i}' = P_i^\gamma$ for each pixel \boldmath$P_i$).}  In rows E and F, $\gamma = 0.3$, while in rows G and H, $\gamma = 0.5$.  Additional galaxy samples can be found in Appendix \ref{GalSamp}.}
    \label{fig:TestCollage}
\end{figure*}

\begin{figure*} 
    \centering
    \begin{minipage}{\textwidth}
        \centering
        \includegraphics[width = \textwidth]{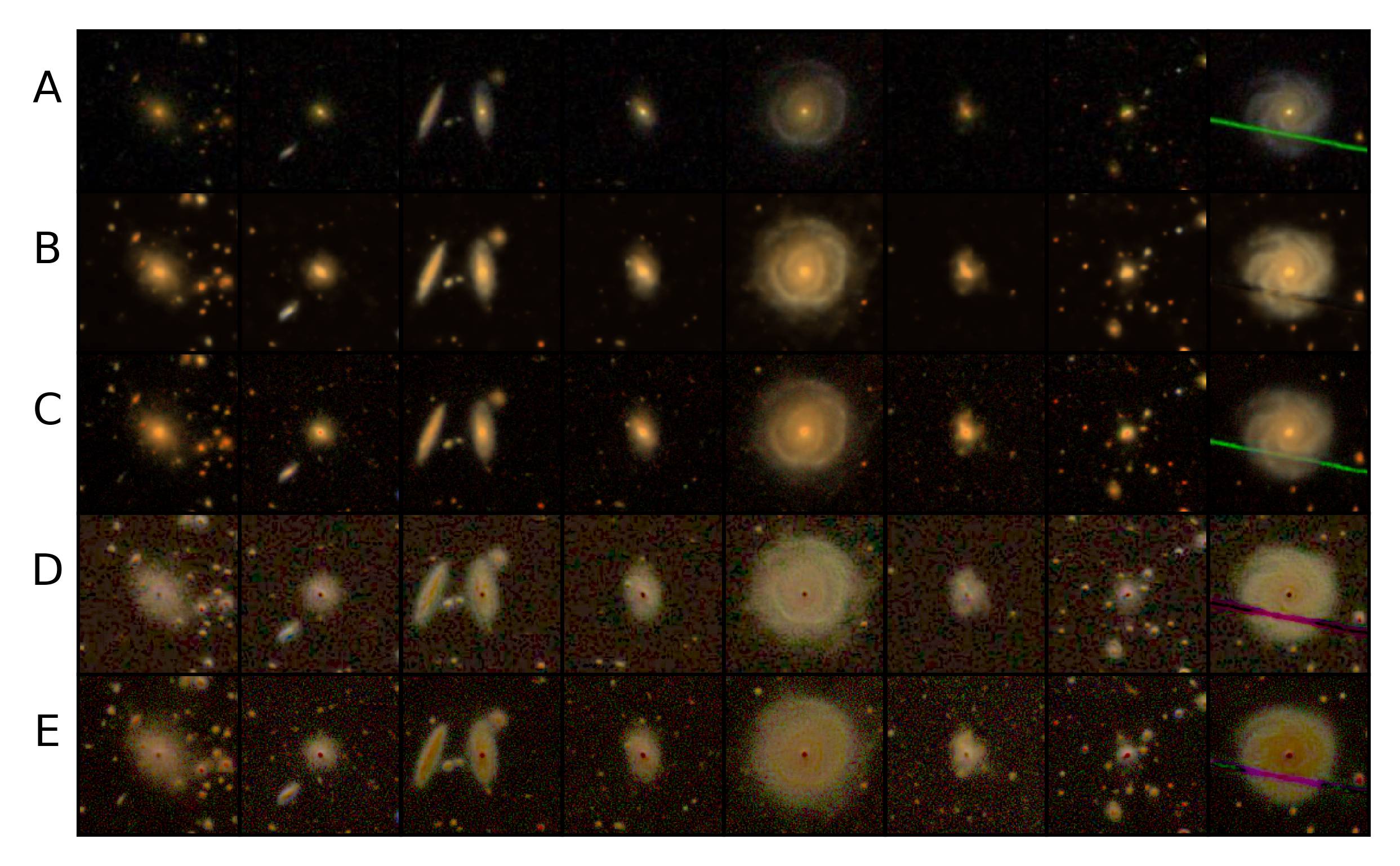}
    \end{minipage}
    \caption{Examples of galaxy images from the external dataset (from outside of the Stripe82 overlap region).  Each column shows an SDSS galaxy (row A) and its DES image reconstruction by the CAE (row B) and CycleGAN (row C) methods.  The CAE and CycleGAN pixel-to-pixel brightness increases (reconstruction - SDSS) are shown in rows D and E, respectively.  \textbf{Note that to increase visibility, images in rows D and E were artificially enhanced with a power law transform (\boldmath$P_i\,\to\,{P_i}' = P_i^\gamma$ for each pixel \boldmath$P_i$, where \boldmath$\gamma = 0.5$).}  Additional galaxy samples can be found in Appendix \ref{GalSamp}.}
    \label{fig:ExtCollage}
\end{figure*}

Fig. \ref{fig:TestCollage} shows several examples of false images generated by the neural networks paired with their corresponding SDSS and DES images from the overlap region.  These images were selected to demonstrate the wide variety of galaxy types and structures included in the validation sample which were not including during the training.  Row A contains images from the SDSS catalog; the corresponding DES images are located in row B.  Rows C and D contain the reconstructed CAE and CycleGAN images, respectively.  We can observe that the DES images and the synthetic images in rows C and D are remarkably similar, where the small differences come from the the lack of structure resolution of the reconstructed objects.

Qualitatively, the reconstructed images are generally blurrier than the corresponding DES images.  However, images reconstructed by both the models are generally brighter than their SDSS counterparts.  In addition, the false images, particularly the those generated by the CAE, are often less noisy than their SDSS and/or DES counterparts.  Image residuals for the CAE and CycleGAN reconstructions are shown in rows E and F, respectively.  These show the pixel-to-pixel brightness differences between the reconstructed and DES images; note that these images were artificially enhanced using a power law transform ($P_i\,\to\,{P_i}' = P_i^\gamma$ for each pixel $P_i$; in rows E and F, $\gamma = 0.3$, while in rows G and H, $\gamma = 0.5$).  This was done so that the residual structure was visible.  It appears that both neural networks isolated and enhanced the galaxy signal while affecting the background minimally or try to reduce the noise.  Both networks were also able to distinguish between separate structures in each image; this is particularly evident in the second column.

Rows G and H show the pixel-to-pixel brightness increase provided by the CAE and CycleGAN reconstructions relative to the corresponding SDSS galaxies, respectively.  Qualitatively, the CAE reconstructions are brighter than the CycleGAN images, and provided greater amplification to the internal structure of each galaxy.  Interestingly, both networks consistently amplified the galaxy center more than other regions.  This amplification was not exclusive to the central galaxy; rather, it was present in most regions the network identified as a signal region.  Other example galaxies are included in Appendix \ref{GalSamp}.

Figure \ref{fig:ExtCollage} shows examples of images from the external dataset (from outside of the Stripe82 region).  These images are generally of lower quality; however, both reconstruction models succeeded in selecting and amplifying the objects of interest with little effect on the background, even maintaining much of the small-scale detail of the images (particularly in the fourth and seventh columns).  As in Fig. \ref{fig:TestCollage}, the reconstructions generally increased the spatial extent of objects in the image.  Notably, the CAE reconstruction appears to have removed an artifact from the SDSS image in the final column; this phenomenon is discussed in greater detail in Section \ref{ImQual}.

\subsection{Dataset Properties} \label{BaseProp}

Here, we quantify the brightness and quality of images from each dataset to use as baseline comparison metrics between the original input SDSS images, the DES target images, and the reconstructions.

\subsubsection*{Pseudo-Flux Magnitude}

In this work we have used the RGB images from SDSS and DES to test our architectures.  Image brightness was quantified using the average pseudo-flux magnitude $F$ of each image.  We refer to $F$ as the ``pseudo-flux magnitude'' because, while $F$ does not represent the physical flux magnitude (our images consisted solely of (r, g, b) channel pixel values), it acts as a proxy for this quantity due to the similarities between the two measurements.  The pseudo-flux magnitude $F$ was defined by

\begin{align} \label{eq:FluxMag}
    F &= 30 - 2.5\log\left(\hspace{-4pt}\sum_{{\hspace{6pt}r_i\,<\,r_{\textnormal{max}}}}\hspace{-10pt} \beta_i\hspace{2pt}\right) \\
    &= 30 - 2.5\log\,\beta^\circ. \nonumber
\end{align}

Here, the pixel brightness $\beta_i$ describes the average of the red, green, and blue channel values in $P_i$ and $\beta^\circ$ is the total pixel brightness contained within an aperture of radius $r_{\textnormal{max}} = 75\pix$.  A constant factor (zero point) of 30 was added to approximate the appearance of a physical magnitude distribution.

\begin{figure*} 
    \centering
    \begin{minipage}{\textwidth}
        \centering
        \includegraphics[width = \textwidth]{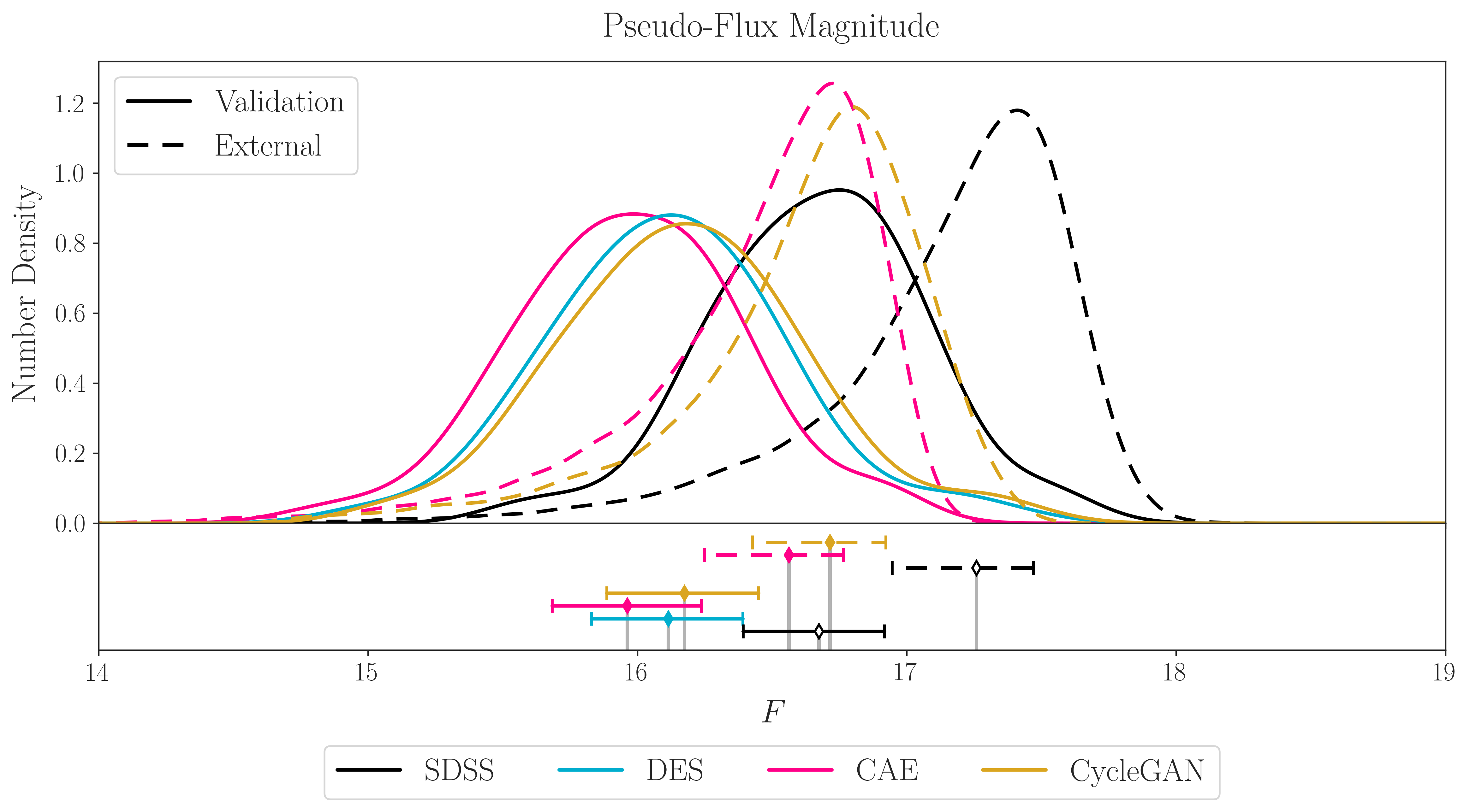}
    \end{minipage}
    \caption{Top: Pseudo-flux magnitudes (defined in Eqn. \eqref{eq:FluxMag}) for the validation and external data.  Bottom:  The first, second, and third quartiles of the corresponding datasets, providing a measure of spread.  SDSS images tended to be fainter than the DES and reconstructed images.  Note that the quartiles cannot be used as a measure of error/statistical significance because this plot does not provide a representation of the image-to-image differences in $F$; this is discussed in greater detail in the text.}
    \label{fig:FluxMag}
\end{figure*}

Gaussian kernel density estimates (KDEs) for histograms of the pseudo-flux magnitudes are shown in Fig. \ref{fig:FluxMag}.  {The first, second, and third quartiles were used as a conservative estimate of the spread of the distribution data; this was chosen due to the heavy skew of the distributions of the external data.  However, they \textit{cannot} be used to determine whether there was a significant difference between the S/N of different datasets.  This is because the differences in pseudo-flux magnitude must be evaluated on an \textit{image-to-image} basis, not by the relative frequency of each S/N value.  The pseudo-flux magnitude values for the reconstructions in the validation datasets were comparable to those of the DES dataset, showing an improvement in the brightness relative to the SDSS dataset.  In the external dataset, the pseudo-flux magnitude distributions for both reconstructions were shifted to the left of the SDSS distribution, indicating that the reconstruction process successfully increased the brightness of images from the external dataset.}

{To quantify the image-to-image brightnesses and provide an error estimate, define the mean relative flux difference $\Delta\Woline[0.85]{\mathcal{F}}^{\kern1pt ij}$ between datasets $i$ and $j$ as}

\begin{align} \label{eq:FluxDiff}
    \RelFluxDiff = \frac{1}{N_{\textnormal{img}}}\sum_m^{N_{\textnormal{img}}}\frac{F_m^{\kern1pt j} - F_m^{\kern1pt i}}{F_m^{\kern1pt i}},
\end{align}

{where $F_m^{\kern1pt i}$ and $F_m^{\kern1pt j}$ are the pseudo-flux magnitudes of corresponding images from the $N_{\textnormal{img}}$ image in datasets $\mathbb{X}_i$ and $\mathbb{X}_j$, respectively.}

{The values of \RelFluxDiff for the external and validation datasets are shown in Table \ref{tab:FluxRat} (scaled by a factor of $10^3$ to enhance readability); the error was estimated using the standard deviation of $\frac{F_m^{\kern1pt j} - F_m^{\kern1pt i}}{F_m^{\kern1pt i}}$.  Quartiles were used to estimate the error of the pseudo-flux magnitude plot (Figure \ref{fig:FluxMag}) due to the clear skew of the data, so the variance would not provide an adequate representation of the spread of the data.  However, the distribution of \RelFluxDiff was more symmetric than those of $F$, allowing the use of the standard deviation as an estimate of error.}

{The only dataset pair in which there was not a significant difference between the fluxes was for CycleGAN vs. DES.  This implies that the CycleGAN reconstructions decreased the flux of the SDSS images to match that of the DES images.  It should be noted that the results from this table seem to be in conflict with those from Figure \ref{fig:FluxMag}.  This is not unexpected: Figure \ref{fig:FluxMag} show the distribution of the relative frequency of \textit{individual} pseudo-flux magnitudes, while the values in Table \ref{tab:FluxRat} show an \textit{image-to-image} comparison of the pseudo-flux magnitudes.  Hence, the values of Table \ref{tab:FluxRat} provide an appropriate measure of the differences in the pseudo-flux magnitudes of the reconstructions relative to their source images.}

{Notably, there was not a statistically significant difference between the values of \RelFluxDiff for $\mathbb{X}_i, \mathbb{X}_j = \textnormal{SDSS, CAE}$ in the validation and external datasets; the same is also true for $\mathbb{X}_i, \mathbb{X}_j = \textnormal{SDSS, CycleGAN}$.  Hence, the decrease in flux provided by the reconstructions were similar for both the validation and external datasets.  This provides evidence for the robustness of our method: the increase in the brightnesses of the false images was the same regardless of the brightnesses of the input images.}

\begin{table}
    \centering
    \setlength\tabcolsep{0pt}
    \setlength\extrarowheight{0pt}
    \begin{tabularx}{\columnwidth}{@{}|C|C|C|C|@{}}
        \hline
        \rule{0pt}{\the\baselineskip}$\RelFluxDiff \times 10^3$ &
        External                         & \multicolumn{2}{|c|}{Validation} \\[2pt]
        \hline
        \diagbox[width=\dimexpr0.25\columnwidth\relax]{
            \vspace{1pt}\hspace{4pt}$\mathbb{X}_j$}{
            \vspace{-6pt}$\mathbb{X}_i$\hspace{6pt}} &
            \vspace{-7pt}SDSS            & \vspace{-7pt}SDSS   & \vspace{-7pt}DES   \\
        \hline
        \rule{0pt}{\dimexpr\baselineskip-1pt\relax}
        {CAE}    & {$-40.65 \pm 5.59$} & {$-41.90 \pm 5.54$} & {$-9.36 \pm 6.72$} \\
        {CycleGAN} & {$-30.44 \pm 7.28$} & {$-29.33 \pm 6.28$} & {$\mathit{3.65 \pm 8.22}$} \\
        {DES}      & {N/A}               & {$-32.80 \pm 9.13$} & {N/A}   \\
        \hline
    \end{tabularx}
    \caption{The mean proportional difference \RelFluxDiff (scaled by a factor of $10^3$) in the pseudo-flux magnitudes (defined in Eq. \eqref{eq:FluxDiff}) between each of the image sets; the standard deviation was used to estimate the error.  The only dataset pair that does not show a significant difference in $F$ is CycleGAN vs. DES.}
    \label{tab:FluxRat}
\end{table}

\begin{figure*} 
    \centering
    \begin{minipage}{\textwidth}
        \centering
        \includegraphics[width = \textwidth]{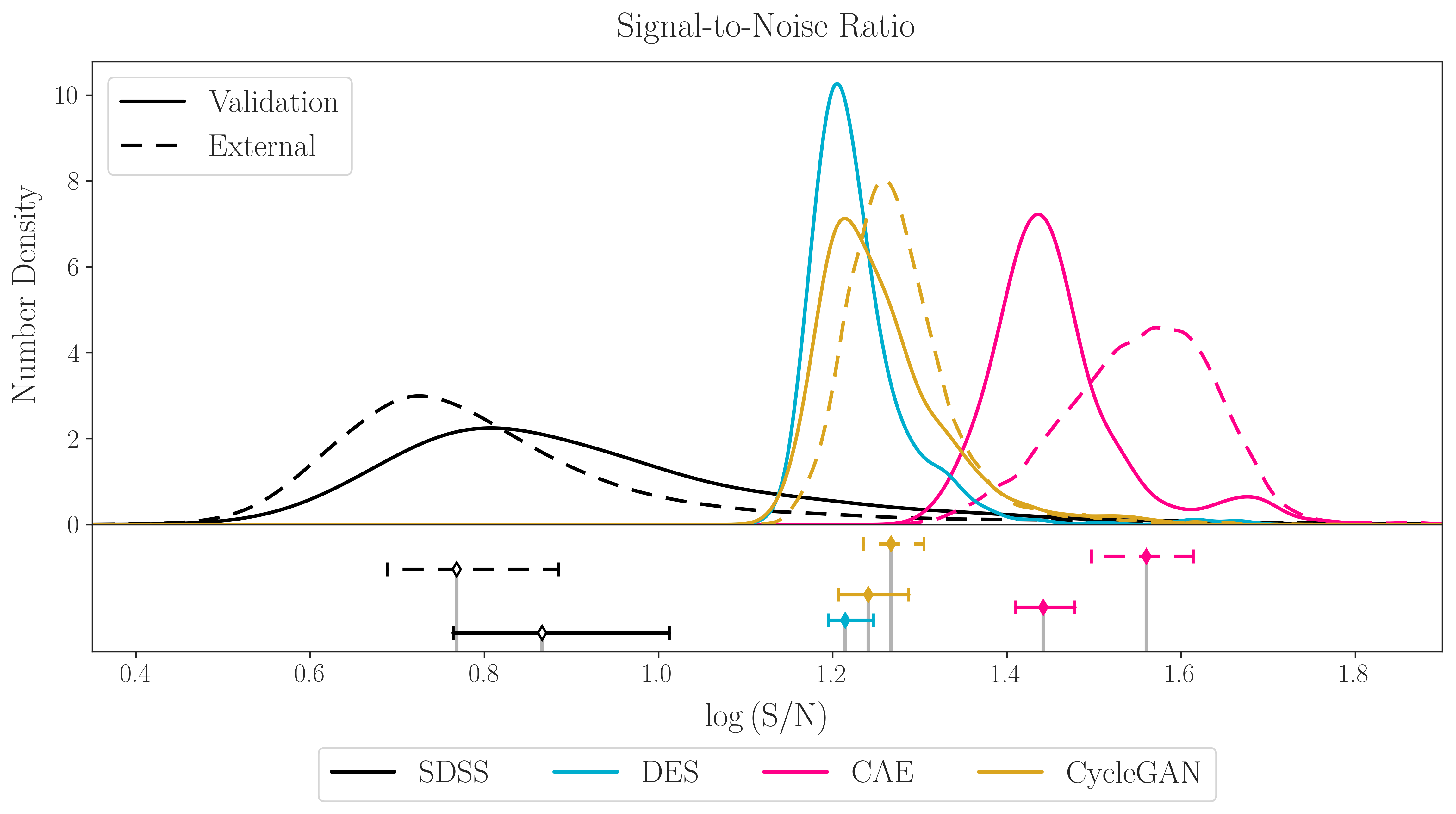}
    \end{minipage}
    \caption{Top: KDEs of the histograms of the mean S/N (defined in Eqn. \eqref{eq:SNR}) for the validation and external data.  Bottom: The first, second, and third quartiles of the distributions.  Both reconstruction models were effective at increasing the S/N.  Note that, as described in Figure \ref{fig:FluxMag}, the quartile bars cannot be used to determine statistical significance; see the text for further explanation.}
    \label{fig:SNR}
\end{figure*}


\subsubsection*{Signal-to-Noise Ratio}

As metric for image quality, we measured the average signal-to-noise ratio (S/N) of images in each dataset.  We define the mean S/N as

\begin{align} \label{eq:SNR}
    \textnormal{S/N} &= \frac{\mu^\circ_\beta}{\sigma^\circ_\beta},
\end{align}

where $\mu^\circ_\beta$ $\left(\sigma^\circ_\beta\right)$ is the mean (standard deviation) of the pixel brightness $\beta$ for pixels within a radius of $r_{\textnormal{max}} = 75\pix$.

{In Fig. \ref{fig:SNR}, we show KDEs for histograms of the mean S/N, along with the first, second, and third quartiles, which are used as a measure of spread; however, as discussed in the analysis of Table \ref{tab:FluxRat}, they cannot be used as a measure of error.  On average, both reconstruction models were effective at boosting the S/N relative to the SDSS images, and the S/N for the CycleGAN reconstructions nearly matched that of the DES images.  Denoising autoencoders have been used to reduce the amount of noise in images \citep{DenoisingAE}, so it is not surprising that the S/N in CAE images was greater than that of the target images.}

{In Table \ref{tab:SNRRat}, we list the mean proportional differences in the signal-to-noise ratios between image sets $i$ and $j$ to summarize the results from Figure \ref{fig:SNR}.  As in Eq. \eqref{eq:FluxDiff}, we define the mean proportional difference between image set $i$ and $j$ as}

\begin{align} \label{eq:SNRDiff}
    \RelSNRDiff = \frac{1}{N_{\textnormal{img}}}\sum_m^{N_{\textnormal{img}}}\frac{\left[\textnormal{S/N}\right]_m^{\kern1pt j} - \left[\textnormal{S/N}\right]_m^{\kern1pt i}}{\left[\textnormal{S/N}\right]_m^{\kern1pt i}}
\end{align}

{where $\left[\textnormal{S/N}\right]_m^{\kern1pt i}$ and $\left[\textnormal{S/N}\right]_m^{\kern1pt j}$ are the signal-to-noise ratios (as defined in Eqn. \eqref{eq:SNR}) of corresponding image pairs from among the $N_{\textnormal{img}}$ images in datasets $\mathbb{X}_i$ and $\mathbb{X}_j$, respectively.  In Table \ref{tab:SNRRat}, we can see that, for the validation dataset, the CycleGAN reconstructions did not provide a significant increase in the S/N relative to the DES images ($\RelSNRDiff - \sigma < 0 < \RelSNRDiff$ for $\mathbb{X}_i, \mathbb{X}_j = \textnormal{SDSS, CycleGAN}$, where $\sigma$ is the standard deviation of \RelSNRDiff).  However, the CAE reconstructions did provide a significant increase over the DES images.  The second and third columns from the left ($\mathbb{X}_i = \textnormal{SDSS}$ for the external and validation data, respectively) indicate that the reconstructions provided a significant increase in the S/N of their corresponding SDSS images.  Moreover, there was not a significant difference between the value of \RelSNRDiff for $\mathbb{X}_i, \mathbb{X}_j = \textnormal{SDSS, CAE}$ in the validation and external datasets; this relationship is the same for $\mathbb{X}_i, \mathbb{X}_j = \textnormal{SDSS, CycleGAN}$.  This implies that image reconstruction via feature translation using our architectures provides a robust method to generate false galaxy images that share the same S/N as DES galaxies in this study.}

\begin{table}
    \centering
    \setlength\tabcolsep{0pt}
    \setlength\extrarowheight{0pt}
    \begin{tabularx}{\columnwidth}{@{}|C|C|C|C|@{}}
        \hline
        \rule{0pt}{\the\baselineskip}$\Delta\Woline[0.95]{\mathcal{S}}^{\kern1pt ij}$ &
        External                         & \multicolumn{2}{|c|}{Validation} \\[2pt]
        \hline
        \diagbox[width=\dimexpr0.25\columnwidth\relax]{
            \vspace{1pt}\hspace{4pt}$\mathbb{X}_j$}{
            \vspace{-6pt}$\mathbb{X}_i$\hspace{6pt}} &
            \vspace{-7pt}SDSS            & \vspace{-7pt}SDSS   & \vspace{-7pt}DES   \\
        \hline
        \rule{0pt}{\dimexpr\baselineskip-1pt\relax}
        {CAE}    & {$5.138 \pm 2.719$} & {$2.893 \pm 1.819$} & {$0.710 \pm 0.317$} \\
        {CycleGAN} & {$2.067 \pm 1.001$} & {$1.334 \pm 0.716$} & {$\mathit{0.069 \pm 0.125}$} \\
        {DES}      & {N/A}               & {$1.224 \pm 0.779$} & {N/A}   \\
        \hline
    \end{tabularx}
    \caption{The mean proportional difference in the signal-to-noise ratios (see Eqn. \eqref{eq:SNRDiff}) between each of the image sets; the standard deviation was used to estimate the error.  As in Table \ref{tab:FluxRat}, the only dataset pair that does not show a significant difference in S/N is CycleGAN vs. DES.}
    \label{tab:SNRRat}
\end{table}


\subsection{Pseudo-Luminosity Profile} \label{StrucAnalysis}


\begin{figure*} 
    \centering
    \begin{minipage}{\textwidth}
        \centering
        \includegraphics[width = \textwidth]{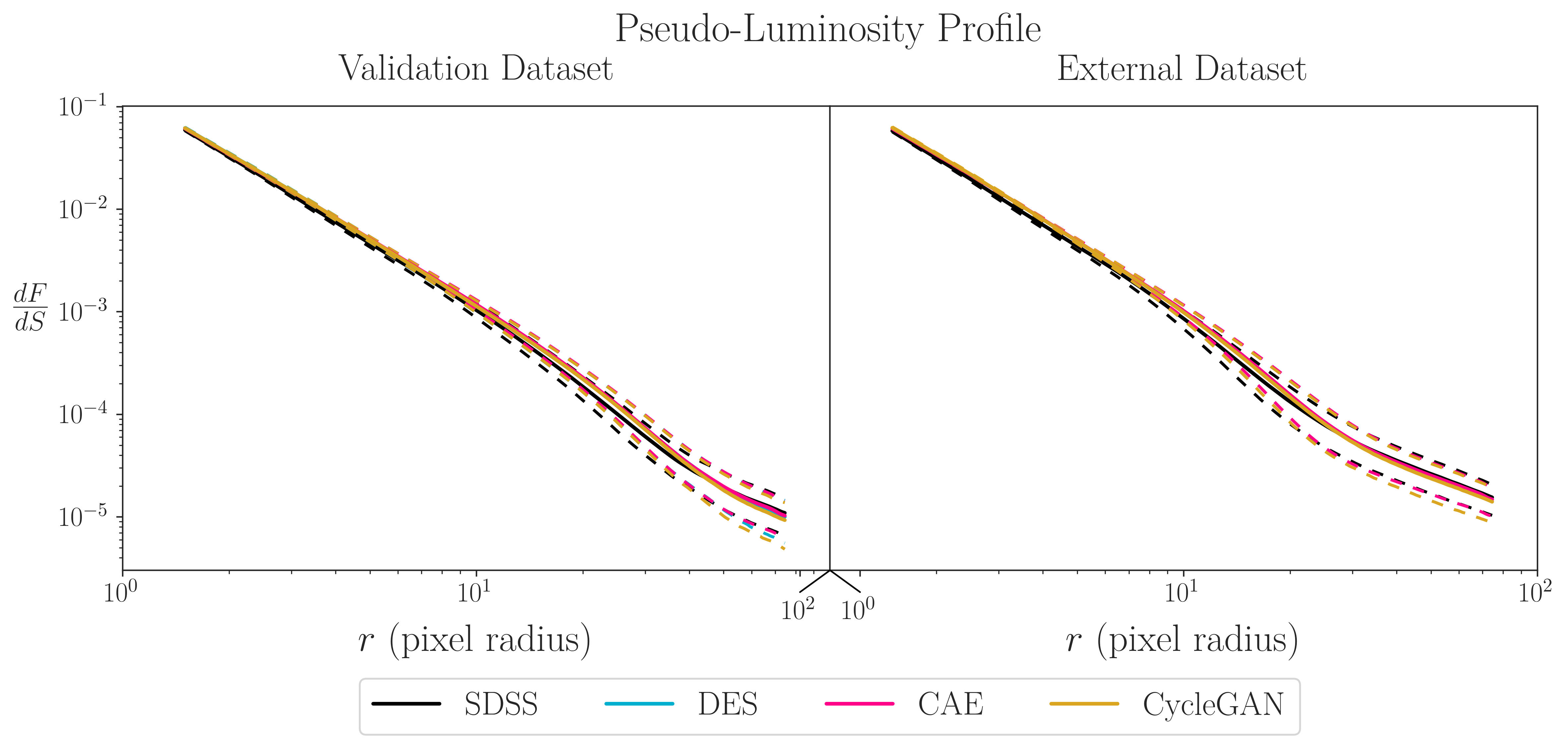}
    \end{minipage}
    \caption{Pseudo-luminosity profiles $\D[F]{S}$ for the validation (left) and external (right) data; $\D[F]{S}$ is defined in Eqn. \eqref{eq:LumProf}.  The solid line represents $\D[F]{S}$, while the dotted lines show $\D[F]{S} \pm \hat{\sigma}$, where $\hat{\sigma}$ is the sample standard deviation.  There was no statistically significant difference between the pseudo-luminosity profiles for any dataset.}
    \label{fig:LumProf}
\end{figure*}

In Section \ref{BaseProp}, we showed that the image reconstructions provided a significant increase in the pseudo-flux magnitude and S/N of their corresponding SDSS images that matched that of the DES images.  We also demonstrate that the improvement in $F$ and S/N were not heavily dependent on the dataset from which the source image was taken, providing evidence for the robustness of our method.  Now, we will compare the pseudo-luminosity profiles of the objects in these images to characterize the structure of the objects themselves.

The pseudo-luminosity profile $\D[F]{S}$,  which is analogous to the luminosity profile in observed data, is defined by

\begin{align} \label{eq:LumProf}
    \D[F(r)]{S} &= \frac{1}{2\pi r\Dpix}\hspace{-4pt}\sum_{\hspace{4pt}r_i\in\textnormal{Ann}(r)}\hspace{-8pt}F_i \\
    &= \frac{F_{\textnormal{Ann}(r)}}{2\pi r\Dpix} \nonumber
\end{align}

where $F_{\textnormal{Ann}(r)}$ is the total flux contained within an annulus-shaped aperture $\textnormal{Ann}(r)$ with central radius $r$ and area $S = 2\pi r\Dpix$, where $\Dpix = 1\pix$

Plots of the pseudo-luminosity profile for the validation and external datasets are shown in Fig. \ref{fig:LumProf}.  {Bootstrapping was used to estimate the sample variance $\hat{\sigma}^2$; the dotted lines represent $\D[F]{S} \pm \hat{\sigma}$.  $\hat{\sigma}^2$ was estimated by resampling each dataset 1000 times; for the validation dataset, the sample size was 50, while for the external dataset, the sample size was 2500.  In both the external and validation datasets, there was no significant difference between the pseudo-luminosity profiles of any image datasets, and from the pseudo-flux magnitude results (Figure \ref{fig:FluxMag}), we know that the reconstructions were generally brighter than their SDSS counterparts.  This implies that the reconstructions improved the brightness quality of the SDSS images without losing information about the object's brightness profile distribution.}

\subsection{Image Quality Comparison} \label{ImQual}

\begin{figure*}
    \centering
    \begin{minipage}{\textwidth}
        \centering
        \includegraphics[width = \textwidth]{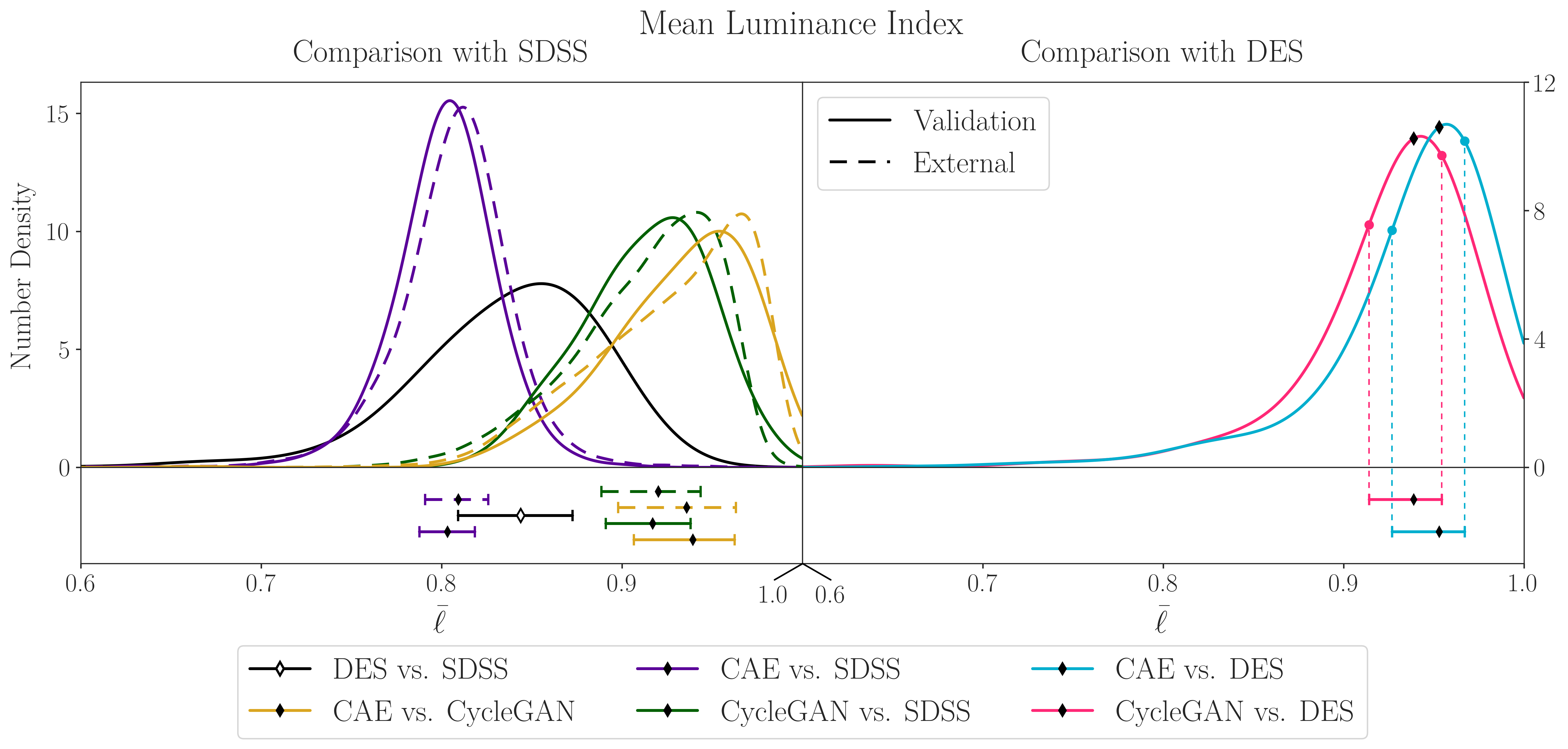}
    \end{minipage}
    \caption{Top: Mean luminance index $\bar{\ell}$ (defined in Eqn. \eqref{eq:lcs}) for the validation and external data.  Bottom: The first, second, and third quartiles of each distribution.  $\bar{\ell}$ describes the similarities in brightness between two images at small scales ($\sim 10 \pix$).  The robustness of the method is indicated by the similarities in the validation and external distributions in the left-hand plot.  In the right-hand plot, both reconstruction models increased $\bar{\ell}$ by a similar amount, indicating that, at small scales, the brightness increase provided by the two models were similar.  This supports the conclusions drawn from the pseudo-flux magnitude in Fig. \ref{fig:FluxMag} and Table \ref{tab:FluxRat}.  Note that unlike in Figures \ref{fig:FluxMag} and \ref{fig:SNR}, the quartiles can be used to determine statistical significance because $\ell$ calculations provide direct image-to-image comparisons.}
    \label{fig:LumId}
\end{figure*}

\begin{figure*}
    \centering
    \begin{minipage}{\textwidth}
        \centering
        \includegraphics[width = \textwidth]{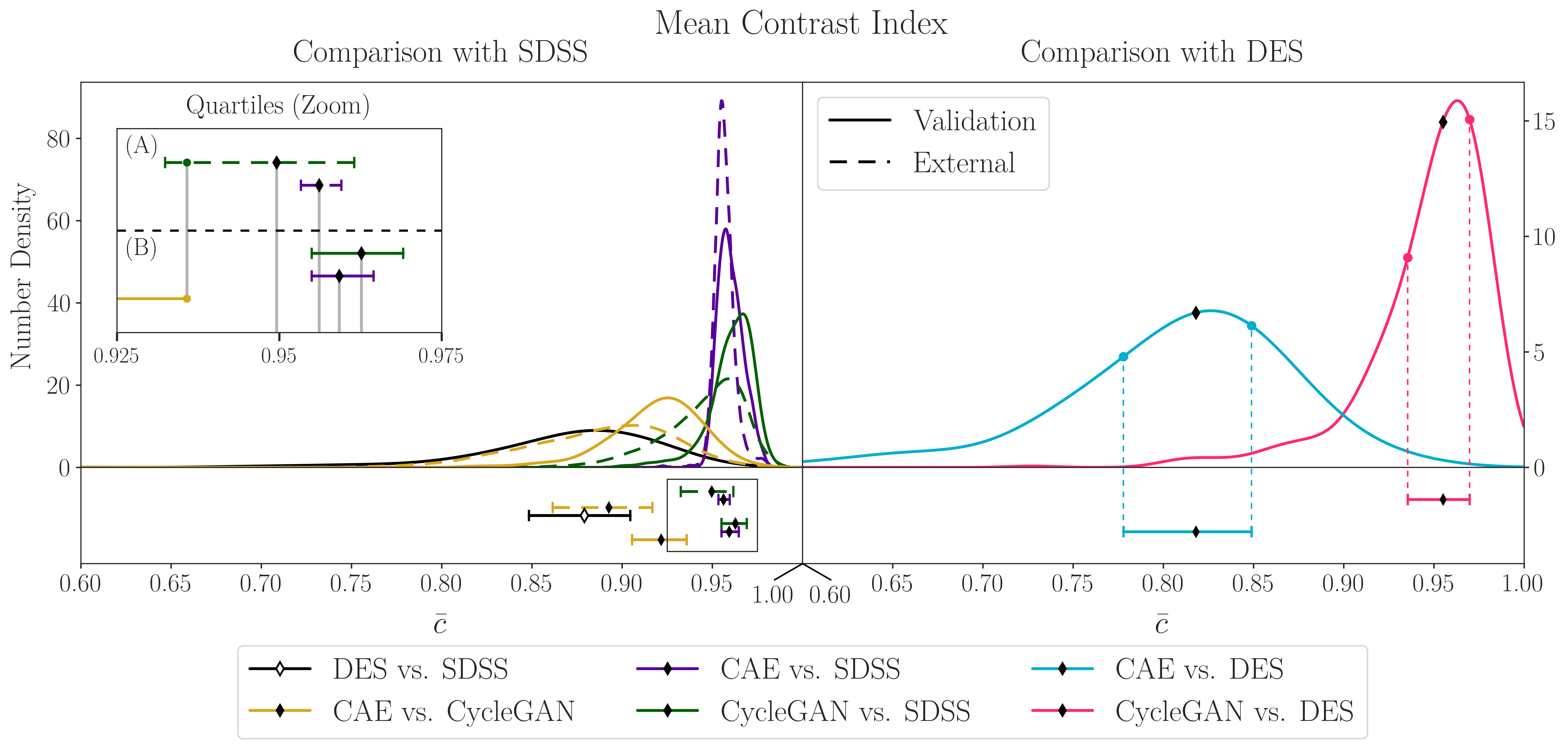}
    \end{minipage}
    \caption{Top: Mean contrast index $\bar{c}$ (defined in Eqn. \eqref{eq:lcs}) for the validation and external data.  Bottom: The first, second, and third quartiles of each distribution.  $\bar{c}$ describes the relative sharpness of two images at small scales ($\sim 10 \pix$).  The robustness of the method is indicated by the similarities in the validation and external distributions in the left-hand plot.  In the right-hand plot, $\bar{c}$ was generally lower for the CAE reconstructions than for the CycleGAN reconstructions, implying that the CAE images were generally blurrier than the CycleGAN images.  This confirms the qualitative observations about the images described in Section \ref{QualAnalysis} (see Fig. \ref{fig:TestCollage}).  Note that unlike in Figures \ref{fig:FluxMag} and \ref{fig:SNR}, the quartiles can be used to determine statistical significance because $c$ calculations provide direct image-to-image comparisons.}
    \label{fig:ContId}
\end{figure*}

\begin{figure*}
    \centering
    \begin{minipage}{\textwidth}
        \centering
        \includegraphics[width = \textwidth]{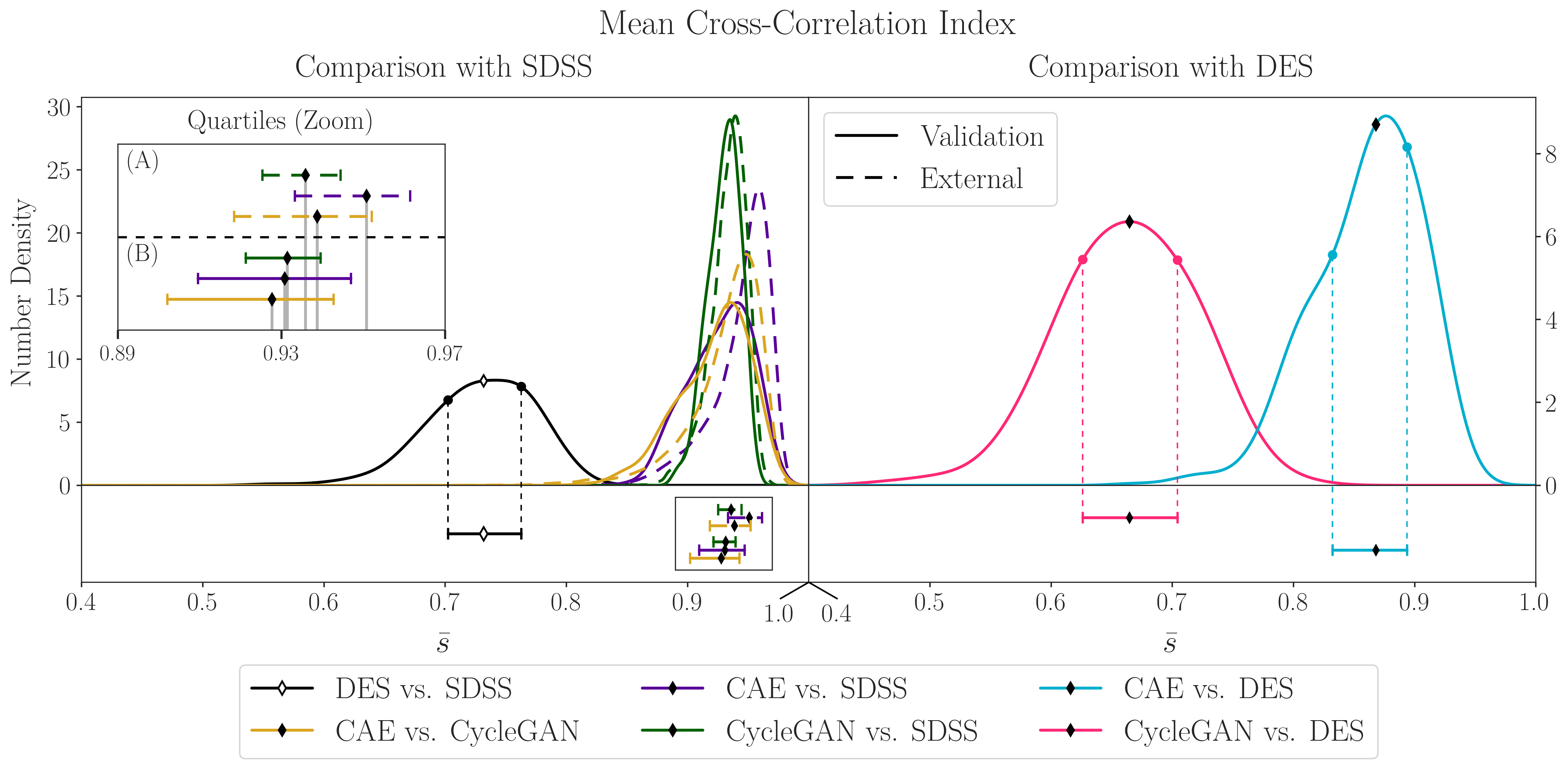}
    \end{minipage}
    \caption{Top: Mean cross-correlation index $\bar{s}$ (defined in Eqn. \eqref{eq:lcs}) for the validation and external data.  Bottom: The first, second, and third quartiles of each distribution.  $\bar{s}$ describes the similarities between the structure of two images at small scales ($\sim 10 \pix$), providing a measure of the faithfulness of the reconstruction.  The robustness of the method is indicated by the similarities in the validation and external distributions in the left-hand plot.  In the right-hand plot, $\bar{s}$ was generally lower for the CycleGAN reconstructions than for the CAE reconstructions, implying that the CAE architecture more accurately recreated small-scale details of the DES images, providing a more accurate reconstruction of the morphological properties of the image.  Note that unlike in Figures \ref{fig:FluxMag} and \ref{fig:SNR}, the quartiles can be used to determine statistical significance because $s$ calculations provide direct image-to-image comparisons.}
    \label{fig:StrucId}
\end{figure*}

\begin{figure*}
    \centering
    \begin{minipage}{\textwidth}
        \centering
        \includegraphics[width = \textwidth]{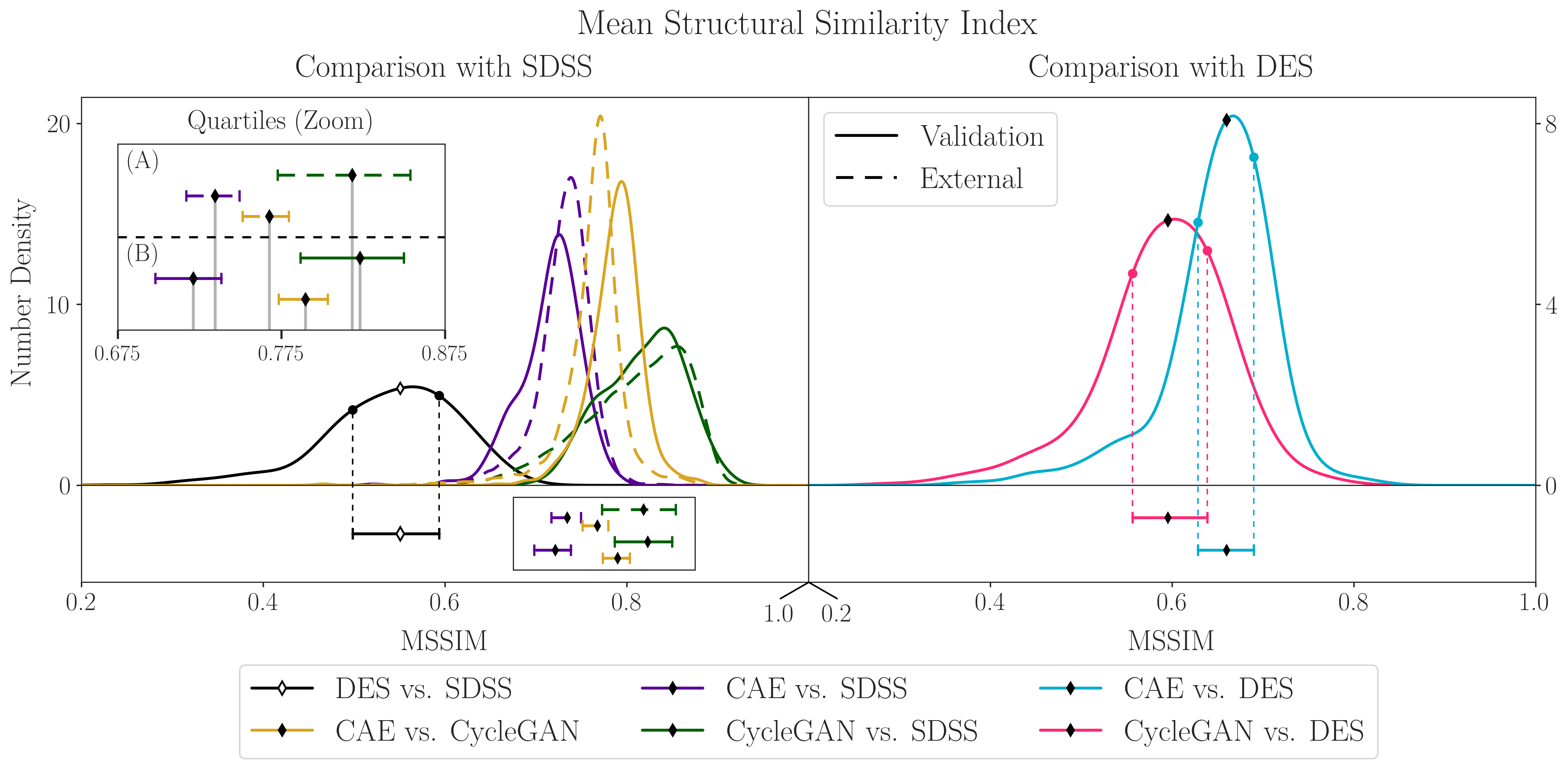}
    \end{minipage}
    \caption{Top: Mean structural similarity index (defined in Eqn. \eqref{eq:MSSIM}) for the validation and external data.  Bottom: The first, second, and third quartiles of each distribution.  The MSSIM, which is the mean of the product of $\ell$, $c$, and $s$, provides a metric for the overall relative image quality.  The robustness of the method is indicated by the similarities in the validation and external distributions in the left-hand plot.  From the right-hand plot, we can see that the overall quality of the CAE images was similar to that of the DES images, while the quality of the CycleGAN reconstructions was further removed from that of the DES images.  Note that unlike in Figures \ref{fig:FluxMag} and \ref{fig:SNR}, the quartiles can be used to determine statistical significance because MSSIM calculations provide direct image-to-image comparisons.}
    \label{fig:SSIM}
\end{figure*}

As shown in Section \ref{BaseProp}, the brightnesses and S/N of the reconstructions were greater than or not significantly different from those of the DES images, and in \ref{StrucAnalysis}, we show that the brightness increase provided by the reconstruction has little effect on the radial profiles of the objects.  Now, we will characterize how effective each reconstruction model is at amplifying the image signal, reducing background noise, improving image quality, and retaining the morphological information contained within the original image.  We also highlight several notable images from the external dataset that show that CAE reconstructions may help remove image artifact.


The mean structural similarity index (MSSIM) \citep{SSIM} is a method used to compare image quality that takes into account differences in brightness, sharpness, and small-scale features.  The MSSIM is defined by the product of the luminance index $\ell$, contrast index $c$, and cross-correlation index $s$.  For a pair of images $\vec{X}$ and $\vec{Y}$, where each respective entry $\vec{X}_{ij}$ and $\vec{Y}_{ij}$ is the pixel brightness $\beta_{ij}$ of pixel $P_{ij}$, let $\vec{x}_{ij}$ $\left(\vec{y}_{ij}\right)$ be an 11\by11 window centered around pixel $x_{ij}$ $\left(y_{ij}\right)$.  After smoothing $\vec{x}_{ij}$ $\left(\vec{y}_{ij}\right)$ by an 11-tap Gaussian filter, define $\ell$, $c$, and $s$ as

\begin{align} \label{eq:lcs}
    \ell\left(\vec{x}_{ij}, \vec{y}_{ij}\right) &= \frac{2\mu_x\mu_y + C_1}{\mu_x^2 + \mu_y^2 + C_1} \nonumber\\
    c\left(\vec{x}_{ij}, \vec{y}_{ij}\right) &= \frac{2\sigma_x\sigma_y + C_2}{\sigma_x^2 + \sigma_y^2 + C_2}, \\
    s\left(\vec{x}_{ij}, \vec{y}_{ij}\right) &= \frac{2\sigma_{xy} + C_2}{2\sigma_x\sigma_y + C_2}. \nonumber
\end{align}

Then structural similarity index SSIM can be calculated as

\begin{align}
    \textnormal{SSIM}\left(\vec{x}_{ij}, \vec{y}_{ij}\right) &= \ell\left(\vec{x}_{ij}, \vec{y}_{ij}\right) c\left(\vec{x}_{ij}, \vec{y}_{ij}\right) s\left(\vec{x}_{ij}, \vec{y}_{ij}\right) \\
    &= \frac{\left(2\mu_x\mu_y + C_1\right) \left(2\sigma_{xy} + C_2\right)}{\left(\mu_x^2 + \mu_y^2 + C_1\right) \left(\sigma_x^2 + \sigma_y^2 + C_2\right)}, \nonumber
\end{align}

Here, $\mu_x$ ($\mu_y$) is the mean of $\vec{x}_{ij}$ $\left(\vec{y}_{ij}\right)$, $\sigma_x^2$ $\left(\sigma_y^2\right)$ is the variance of $\vec{x}_{ij}$ $\left(\vec{y}_{ij}\right)$, $\sigma_{xy}^2$ is the covariance, $C_1 = (0.01 R_D)^2$, and $C_2 = (0.03 R_D)^2$ are stabilization constants for which $R_D$ is the dynamic range of the image (in our case, $R_D = 1$).  Then the MSSIM is defined by

\begin{align} \label{eq:MSSIM}
    \textnormal{MSSIM} = \frac{1}{N_P^2}\sum_{i,j}^{N_P}\textnormal{SSIM}\left(\vec{x}_{i,j},\vec{y}_{i,j}\right)
\end{align}

and the mean luminance, contrast, and cross-correlation indices ($\bar{\ell}$, $\bar{c}$, and $\bar{s}$, respectively) are defined similarly.

KDEs of histograms for $\bar{\ell}$, $\bar{c}$, $\bar{s}$, and MSSIM for the overlap and external data are shown in Figures \ref{fig:LumId}, \ref{fig:ContId}, \ref{fig:StrucId}, and \ref{fig:SSIM}, respectively.

As the SDSS galaxies are substantially different in brightness and radii, it is not valid to use $\bar{\ell}$, $\bar{c}$, $\bar{s}$, and MSSIM as image quality metrics for DES/SDSS and reconstruction/SDSS image pairs.  However, if the reconstruction process is robust, the distributions for DES/SDSS pairs and reconstruction/SDSS pairs should be consistent in the validation and external datasets.  Hence, we will use reconstruction/DES and reconstruction/reconstruction measurements to quantify the reconstruction quality and reconstruction/SDSS measurements as metrics for robustness.

{The mean luminance index $\bar{\ell}$ is a measure of the differences in the pixel-to-pixel brightness of two (smoothed) images.  The reconstruction/DES distributions for $\bar{\ell}$ were similar in shape, and there was not a significant difference between their medians, indicating that they had similar brightness qualities to one another; this is consistent with the pseudo-flux magnitude results (Figure \ref{fig:FluxMag} and Table \ref{tab:FluxRat}).  The brightness quality of the reconstructions relative to their SDSS counterparts were extremely similar to one another in both the validation and external distributions, implying that both were equally effective at increasing the image brightnesses.}

{While the $\bar{\ell}$ values for the external dataset cannot be interpreted as measures of the image brightness qualities, they can be used to support the robustness of the reconstruction process.  The shapes of the CAE/SDSS $\bar{\ell}$ distributions for the validation and external datasets were similar to one another, and there was not a significant difference between their medians; the same is true for CycleGAN/SDSS.  This implies that the brightness quality improvement was consistent for both the validation and external SDSS datasets, providing evidence for the robustness of this method.}

{The mean contrast index $\bar{c}$ describes the average difference in smoothness between small cut-outs of image pairs.  For the validation data, the CycleGAN reconstructions had a significantly higher contrast index than the CAE reconstructions, implying that the sharpness of the CycleGAN images was more consistent with that of the DES images.  This confirms that the CAE reconstructions tended to smooth the images, leading to the blurriness seen in Figures \ref{fig:TestCollage} and \ref{fig:ExtCollage}.  The robustness of the reconstructions can again be seen by the lack of a significant difference between the test and external distributions for the reconstruction/SDSS $\bar{c}$ distributions.  Note that the differences between the S/N for each image set likely contributed to the value of $\bar{c}$; however, the qualitative sharpness of the reconstructions are consistent with the conclusions drawn from $\bar{c}$.}

{The mean cross-correlation index $\bar{s}$ is a measure of the deviations in the small-scale structure between two images; large values of $\bar{s}$ indicate that, after normalizing for the brightness and sharpness, the morphological features of the images at small scales are similar to (strongly correlated with) one another.  The $\bar{s}$ distributions for the validation data indicates that the CAE reconstructions are significantly more closely correlated with their DES counterparts at small scales than the CycleGAN reconstructions.  This implies that CAE reconstruction preserves more information at small scales than CycleGAN.}

The combination of these quantities yields the MSSIM distributions seen in Fig. \ref{fig:SSIM}.  This metric indicates that the overall quality of the CAE images was comparable to that of the CycleGAN reconstructions; however, the breakdown in terms of $\bar{\ell}$, $\bar{c}$, and $\bar{s}$ suggests that the reconstruction methods provide differing benefits.  Specifically, CycleGAN reconstructions are generally sharper than their CAE counterparts, while CAE reconstructions preserve more information at small scales in the image.

\begin{figure*} 
    \centering
    \begin{minipage}{\textwidth}
        \centering
        \includegraphics[width = \textwidth]{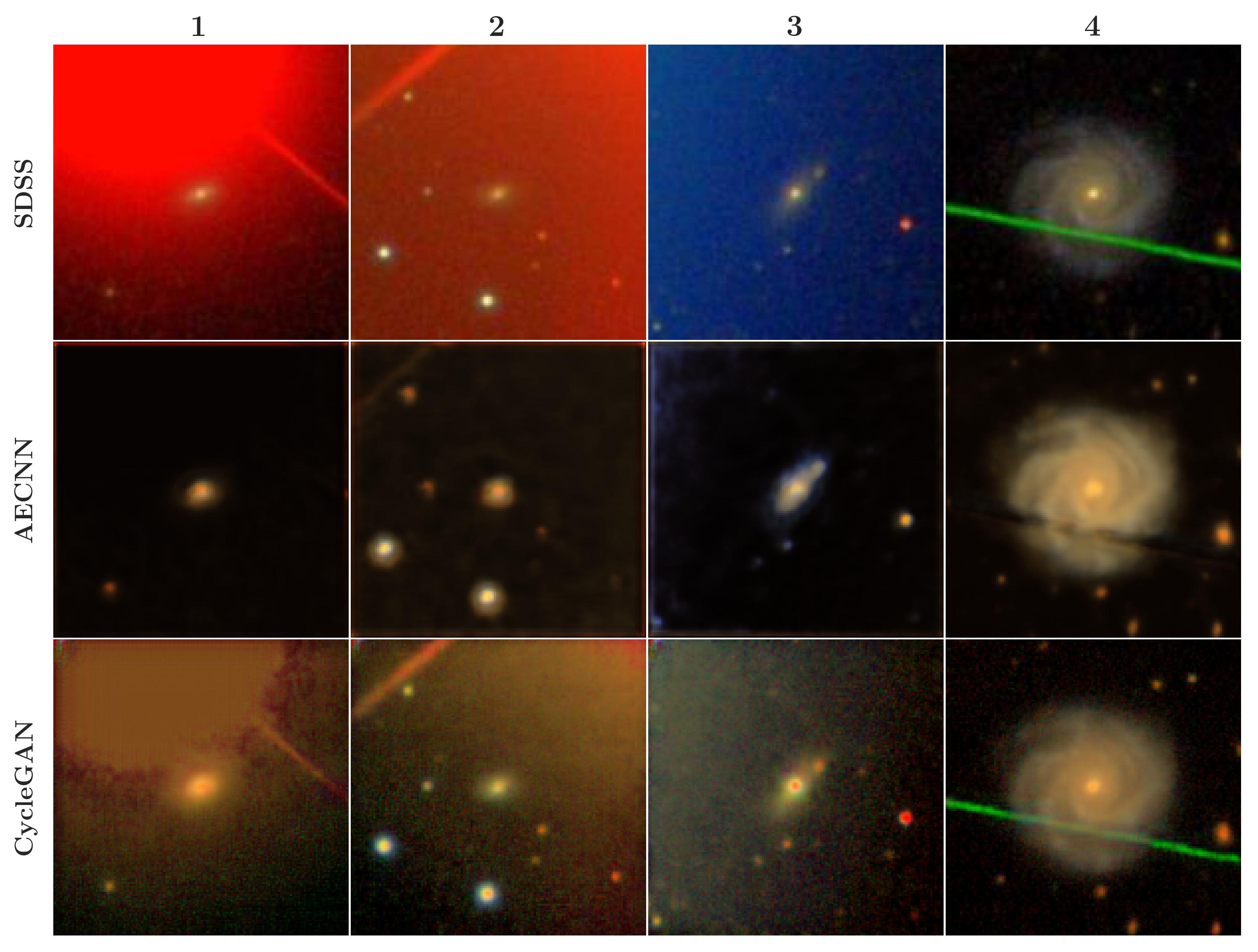}
    \end{minipage}
    \caption{A selection of several notable objects from the external dataset.  In each of these images, it appears that the CAE reconstructions may have removed artifacts from the image.  {The reconstructed objects may have been generated through inpainting.}
    \label{fig:NRemExt}}
\end{figure*}

\begin{figure*} 
    \centering
    \begin{minipage}{\textwidth}
        \centering
        \includegraphics[width = \textwidth]{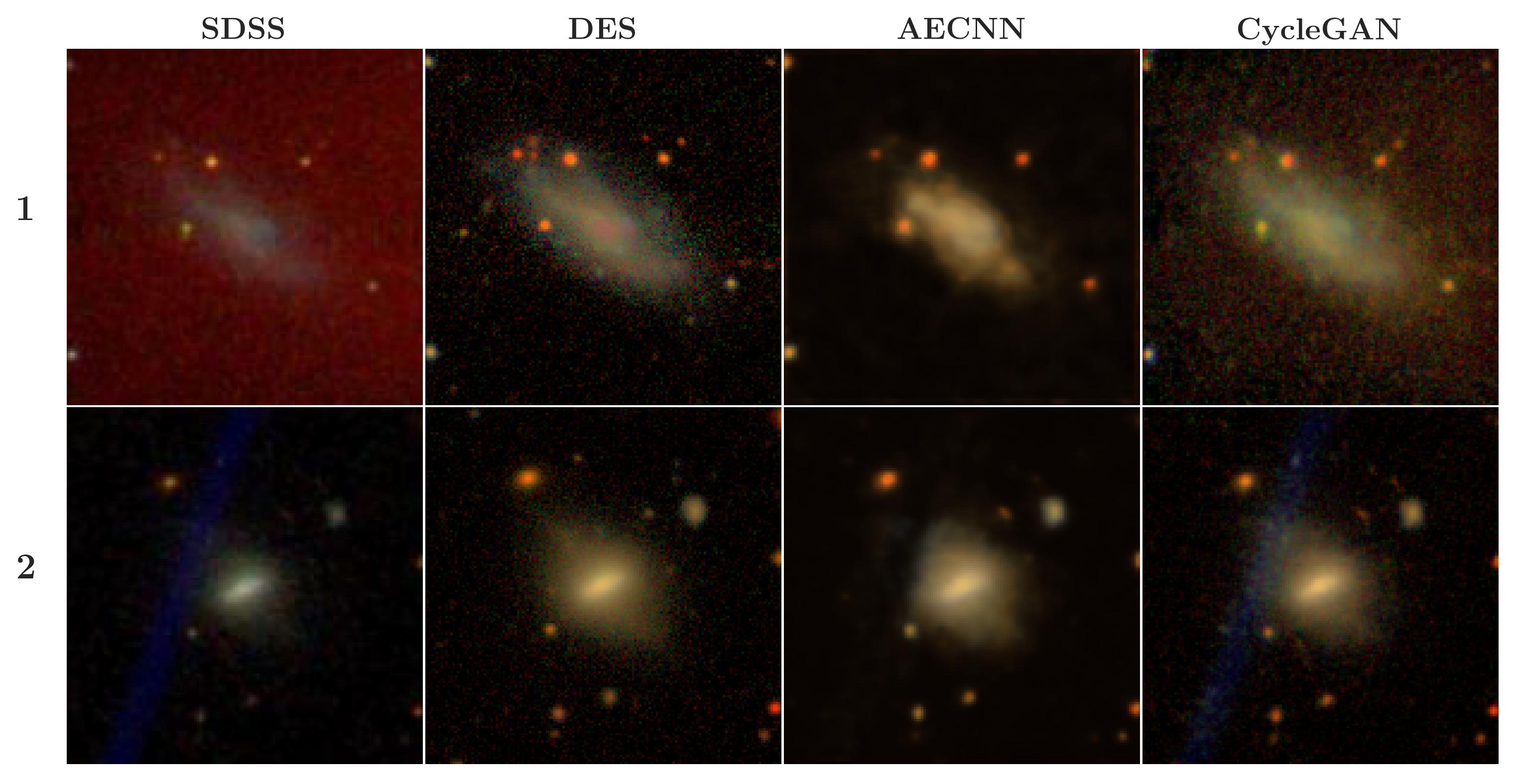}
    \end{minipage}
    \caption{{A selection of several notable objects from the validation dataset.  In each of these images, it appears that the CAE reconstructions may have removed large artifacts from the image.  Note that both reconstructions may have used inpainting to generate the images in column 1, while the CAE reconstruction of the central object in column 2 appears to have removed the corrupted region of that object.}}
    \label{fig:NRemTest}
\end{figure*}

Finally, we would like to highlight several unique images from the external dataset; these are shown in Figure \ref{fig:NRemExt}.  These images were found through visual inspection of images with the lowest reconstruction/SDSS $\bar{\ell}$, $\bar{c}$, $\bar{s}$, and/or MSSIM values in the external dataset.

{Each image in Figure \ref{fig:NRemExt} is heavily corrupted by artifacts; however, the CAE reconstructions appear to have removed these artifacts at the cost of blurring the objects in the image.  These results are consistent with studies of denoising autoencoders \citep{DenoisingAE}, which proven effective at smoothing brightness/color variations, removing artifacts, and restoring corrupted images.  As the base architecture of a denoising autoencoder is similar to that of our encoder/decoder pair, it is not surprising that the image reconstructions were effective at removing these artifacts.  The CycleGAN reconstructions, however, fail to consistently remove these artifacts, though do succeed in amplifying the brightness of these objects.}

{Figure \ref{fig:NRemTest} shows several images from the validation dataset that contain artifacts.  The images in row 1 were found due to their extreme reconstruction/SDSS $\bar{\ell}$, $\bar{c}$, $\bar{s}$, and MSSIM values; however, those in row 2 were found via manual inspection of the validation dataset.  This was to be expected because the Stripe82 dataset is generally of higher quality than the external dataset.}

{In row 1, it appears that the CAE reconstruction removed the artifact, albeit at the cost of blurring the central object.  The artifact in row 2 consists of a blue streak passing through the upper-left edge of the central object.  Like column 4 in Figure \ref{fig:NRemExt}, there is no signal in this region of the CAE reconstruction, implying that little or no inpainting was performed.}

{As the validation data and training data were taken from the same population, it is likely that the training data had a similar incidence of corrupted images as the validation data.  As a result, it is unlikely that either neural network was trained sufficiently to accurately extract the signal from the heavily corrupted images in Figure \ref{fig:NRemExt}, implying that objects recovered in these images likely resulted from inpainting.  While outside the scope of this work, the improvement in the quality of the images in Figure \ref{fig:NRemExt}, especially given the lack of training on corrupted images, warrants a more thorough analysis of the effectiveness of corrupted image reconstruction using our CAE architecture.}

\section{Conclusions} \label{Conclusion}

In this work, we demonstrated the viability of robust cross-survey galaxy image translation using neural networks and generative models.  Using the pseudo-flux magnitude (Section \ref{BaseProp}) and mean luminance index $\bar{\ell}$ (Section \ref{ImQual}), we show that the average brightnesses of the reconstructions more closely match DES images than their SDSS source images while preserving the structural information contained within the source galaxy (Section \ref{StrucAnalysis}).  {In Section \ref{BaseProp}, we also demonstrated that the reconstruction process improved the signal-to-noise ratio of the source images.  The signal-to-noise ratio of the CycleGAN images closely correlated with that of the DES images, while the CAE images improved this quantity relative to the DES images; this behavior is expected because autoencoders have been shown to be effective at reducing the amount of noise in images \citep{DenoisingAE}.}  Together, these imply that our method can be used to improve image brightness and signal strength using image-to-image translation.  In Section \ref{ImQual}, we discuss the pros and cons of each reconstruction method using the mean contrast index $\bar{c}$ and cross-correlation index $\bar{s}$.  We found that CycleGAN reconstructions were sharper, while CAE reconstructions more accurately reproduced the structure of DES galaxies at length scales on the order of several pixels at the cost of being slightly blurrier.  Finally, we highlighted several instances in which the reconstructions appear to have removed large artifacts.  {We find evidence for the robustness of our method by performing reconstructions on images from the SDSS catalog in the external region, which contains objects without a DES counterpart.  Though these images were fainter and had lower S/N than images from the overlap region (Stripe82),  the large- and small-scale statistics of these image reconstructions were similar to those in the overlap region, implying that the reconstruction process accurately created DES representation of these objects.  However, there is the possibility that our model was overfitted due to our choice to avoid factors that may impact the accuracy of the map between SDSS and DES images in the Stripe82 region.}

{While this only constitutes an initial application, our results show that feature transfer learning shows promise as a method for false galaxy image generation.  This has great implications for the analysis of astronomical survey data: assuming that there is a sufficiently large sample of corresponding SDSS and DES image pairs, one could improve the brightness and S/N of many images from the SDSS catalog, decreasing the amount of error and improving the statistical power of analyses.  Additionally, this provides an important advantage over other generative models used supplement survey data: while other methods generate false images that share the properties of the images in the data set of interest, feature-to-feature translation provides representations of observed galaxies, providing a way to extend both the size and the sky coverage of galaxy surveys.}

The reconstruction pipeline we developed solely constitutes a initial exploration, but the efficiency and robustness of the reconstruction process shows promise as a method for generating or improving survey data.  {While SDSS and DES data were used in this work, we expect that this may be applicable to other surveys, particularly for deeper surveys such as LSST \citep{LSST}.  All quantities calculated were derived solely from the mean of the (r, g, b) channel pixel values of survey images; however, we anticipate that similar methods could be used for the generation of false images with physical observables consistent with those of survey images.}  In addition, our methodology could be expanded to enable cross-wavelength or band-to-band translation.  A neural network could be trained with a feature set containing fewer bands than the target dataset, generating a map between each pair of bands in the training and target data.  {The trained network could be used to supplement survey data by generating realistic reconstructions of image data in frequency bands not probed by that survey.  We intend to explore these applications in future work using DES DR2 data, which contains more images and has a greater field depth than DES DR1 \citep{DESDR2}}.

\section{Acknowledgments} 

\noindent This material is based upon work supported by the National Science Foundation Graduate Research Fellowship Program under Grant No. DGE --- 1746047.  M. Carrasco Kind has been supported by NSF Grant AST-1536171.

\subsection*{Author contribution}
B. Buncher: Data analysis, figure creation, writing, and editing. \\
A. N. Sharma: AI model creation, data collection, figure creation, writing, and editing. \\
M. Carrasco Kind: Oversight, data collection, writing, and editing.

\subsection*{Softwares Used}

This research made us of \texttt{matplotlib} \citep{matplotlib}, \texttt{numpy} \citep{numpy1,numpy2}, \texttt{scikit-image} \citep{skimage}, \texttt{SciPY} \citep{SciPy}, and \texttt{seaborn} \citep{seaborn}.

This research made use of \texttt{Astropy},\footnote{http://www.astropy.org} a community-developed core Python package for Astronomy \citep{astropy1,astropy2}.

This research made use of \texttt{Photutils}, an \texttt{Astropy} package for detection and photometry of astronomical sources \citep{photutils}.

\section*{Data Availability Statement}
The data underlying this article will be shared on reasonable request to the corresponding author.

\bibliographystyle{mnras}
\bibliography{main}


\onecolumn
\appendix
\section{Additional Image Samples} \label{GalSamp}

Here, we show additional examples of SDSS, DES, and reconstructed images from the validation dataset similar to Fig. \ref{fig:TestCollage}.  They were randomly selected to provide examples of objects with a variety of types, brightnesses, and extents.

\begin{figure*} 
    \begin{minipage}{\textwidth}
        \centering
        \includegraphics[width = \textwidth]{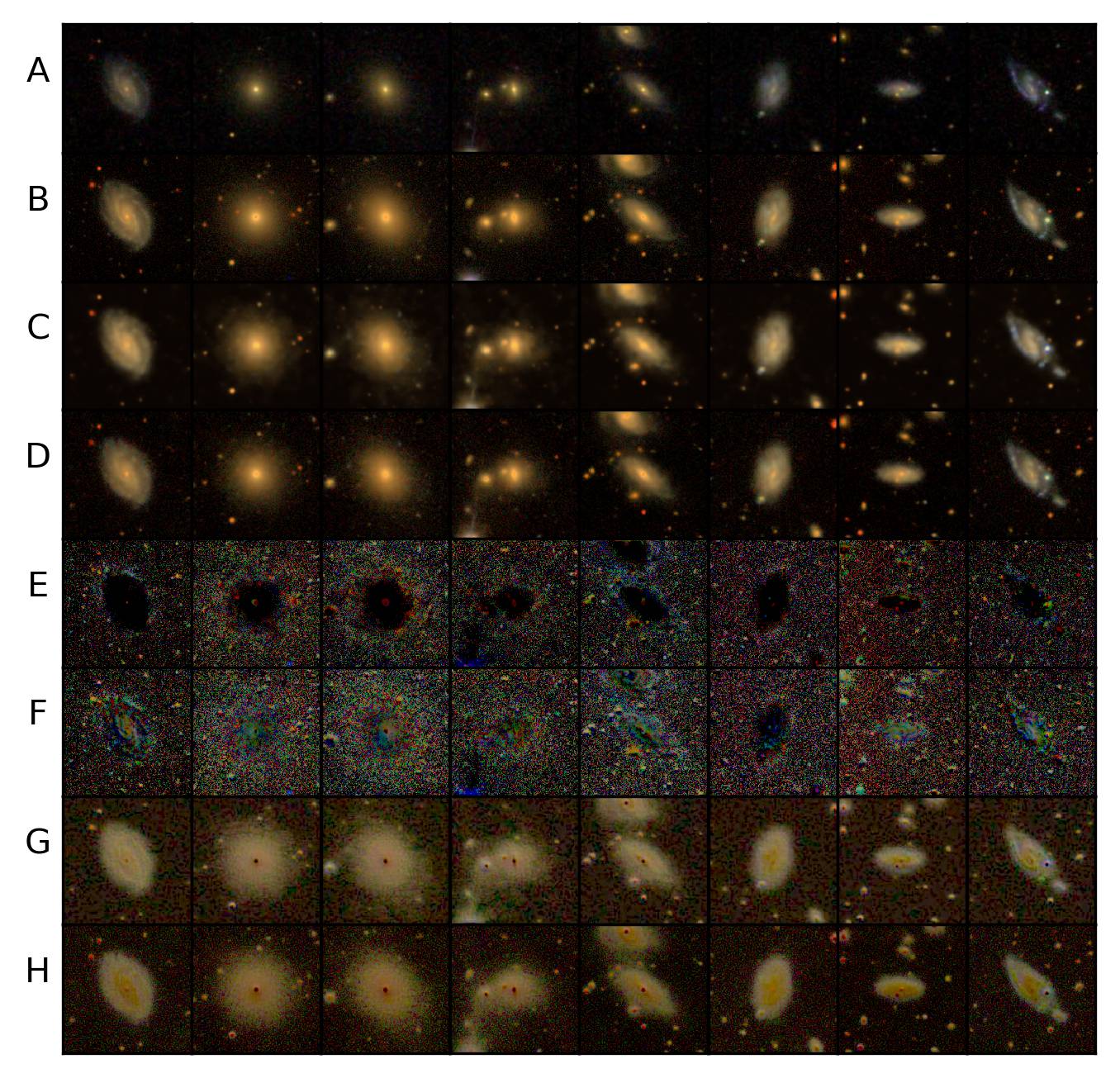}
    \end{minipage}
    \caption{Additional examples of source, target, and reconstructed images from the validation dataset; the formatting is the same as in Figure \ref{fig:TestCollage}.  Note that the images in rows E, F, G and H were enhanced for clarity.}
    \label{fig:Collage1}
\end{figure*}

\begin{figure*} 
    \begin{minipage}{\textwidth}
        \centering
        \includegraphics[width = \textwidth]{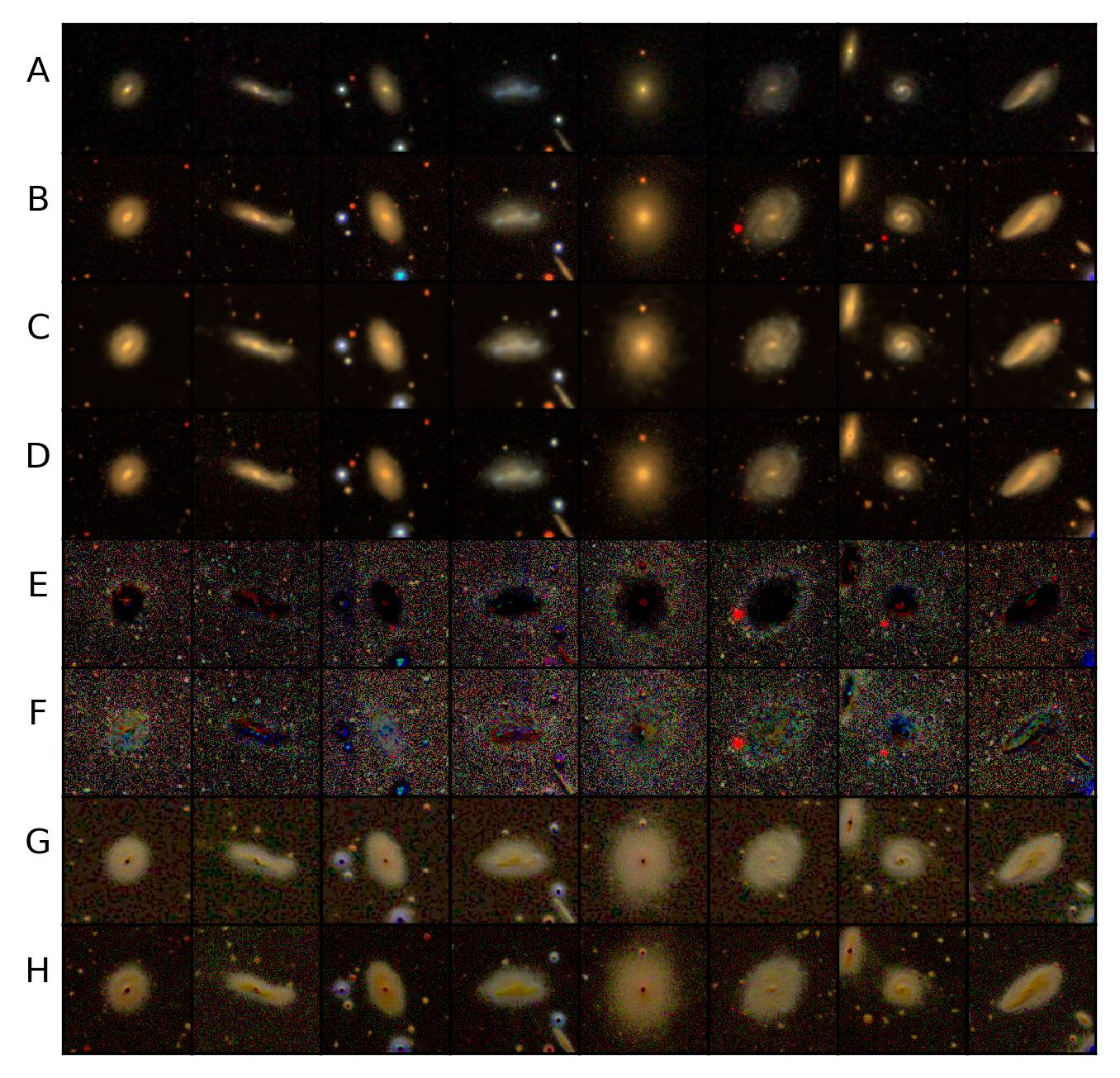}
    \end{minipage}
    \caption{Additional examples of source, target, and reconstructed images from the validation dataset; the formatting is the same as in Figure \ref{fig:TestCollage}.  Note that the images in rows E, F, G and H were enhanced for clarity.}
    \label{fig:Collage2}
\end{figure*}

\begin{figure*} 
    \begin{minipage}{\textwidth}
        \centering
        \includegraphics[width = \textwidth]{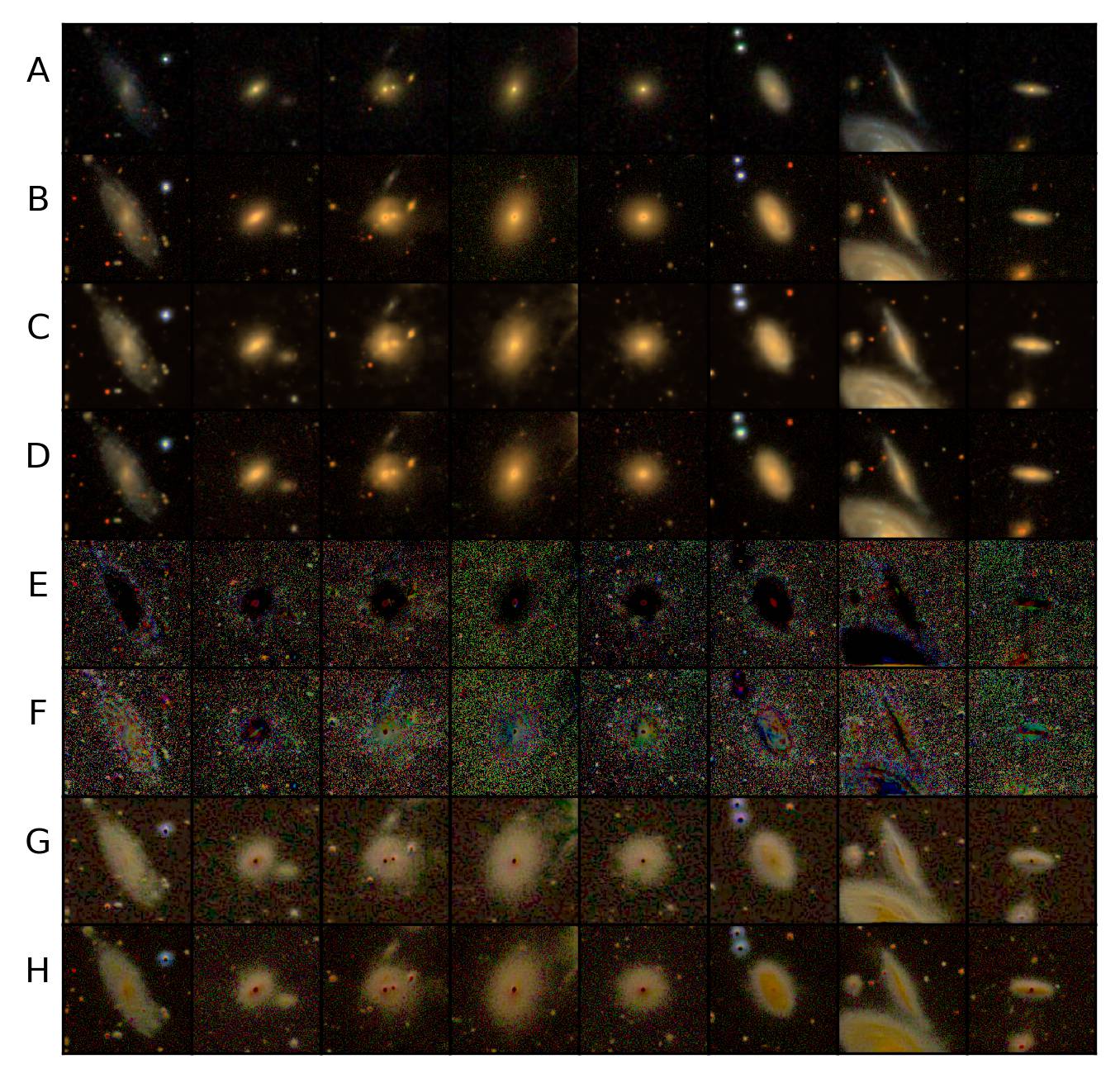}
    \end{minipage}
    \caption{Additional examples of source, target, and reconstructed images from the validation dataset; the formatting is the same as in Figure \ref{fig:TestCollage}.  Note that the images in rows E, F, G and H were enhanced for clarity.}
    \label{fig:Collage3}
\end{figure*}

\begin{figure*} 
    \begin{minipage}{\textwidth}
        \centering
        \includegraphics[width = \textwidth]{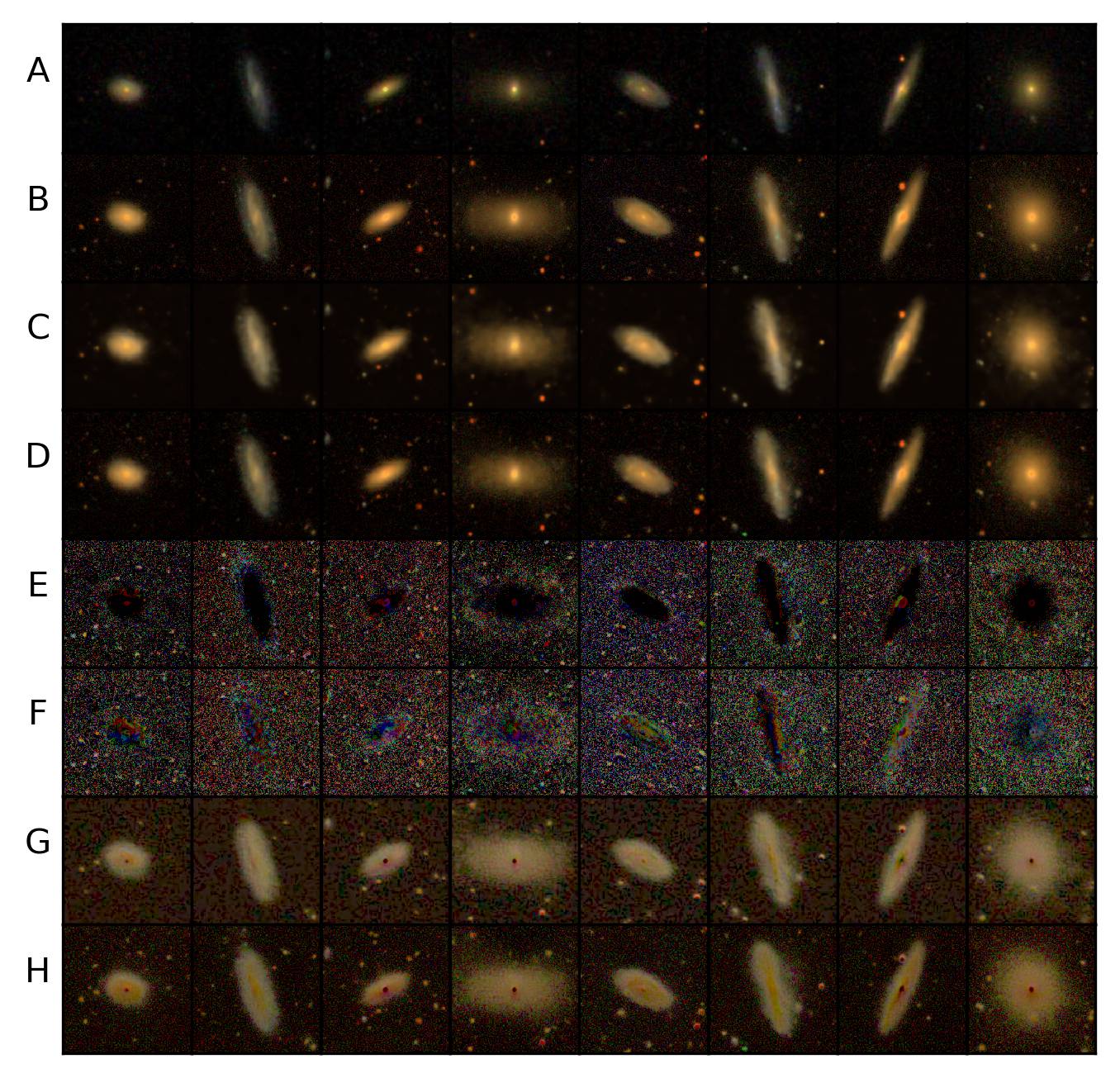}
    \end{minipage}
    \caption{Additional examples of source, target, and reconstructed images from the validation dataset; the formatting is the same as in Figure \ref{fig:TestCollage}.  Note that the images in rows E, F, G and H were enhanced for clarity.}
    \label{fig:Collage4}
\end{figure*}

\begin{figure*} 
    \begin{minipage}{\textwidth}
        \centering
        \includegraphics[width = \textwidth]{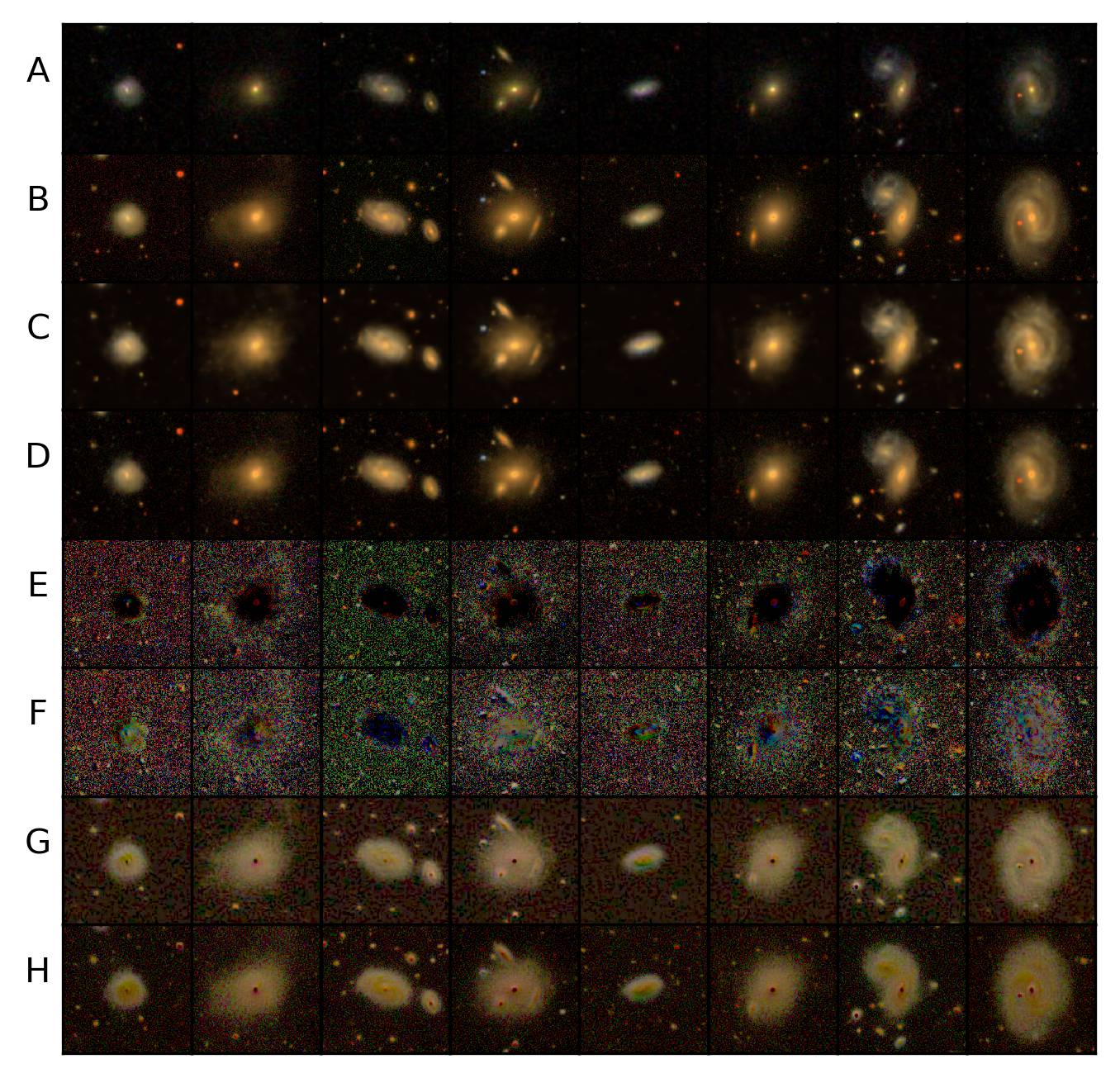}
    \end{minipage}
    \caption{Additional examples of source, target, and reconstructed images from the validation dataset; the formatting is the same as in Figure \ref{fig:TestCollage}.  Note that the images in rows E, F, G and H were enhanced for clarity.}
    \label{fig:Collage6}
\end{figure*}

\begin{figure*} 
    \begin{minipage}{\textwidth}
        \centering
        \includegraphics[width = \textwidth]{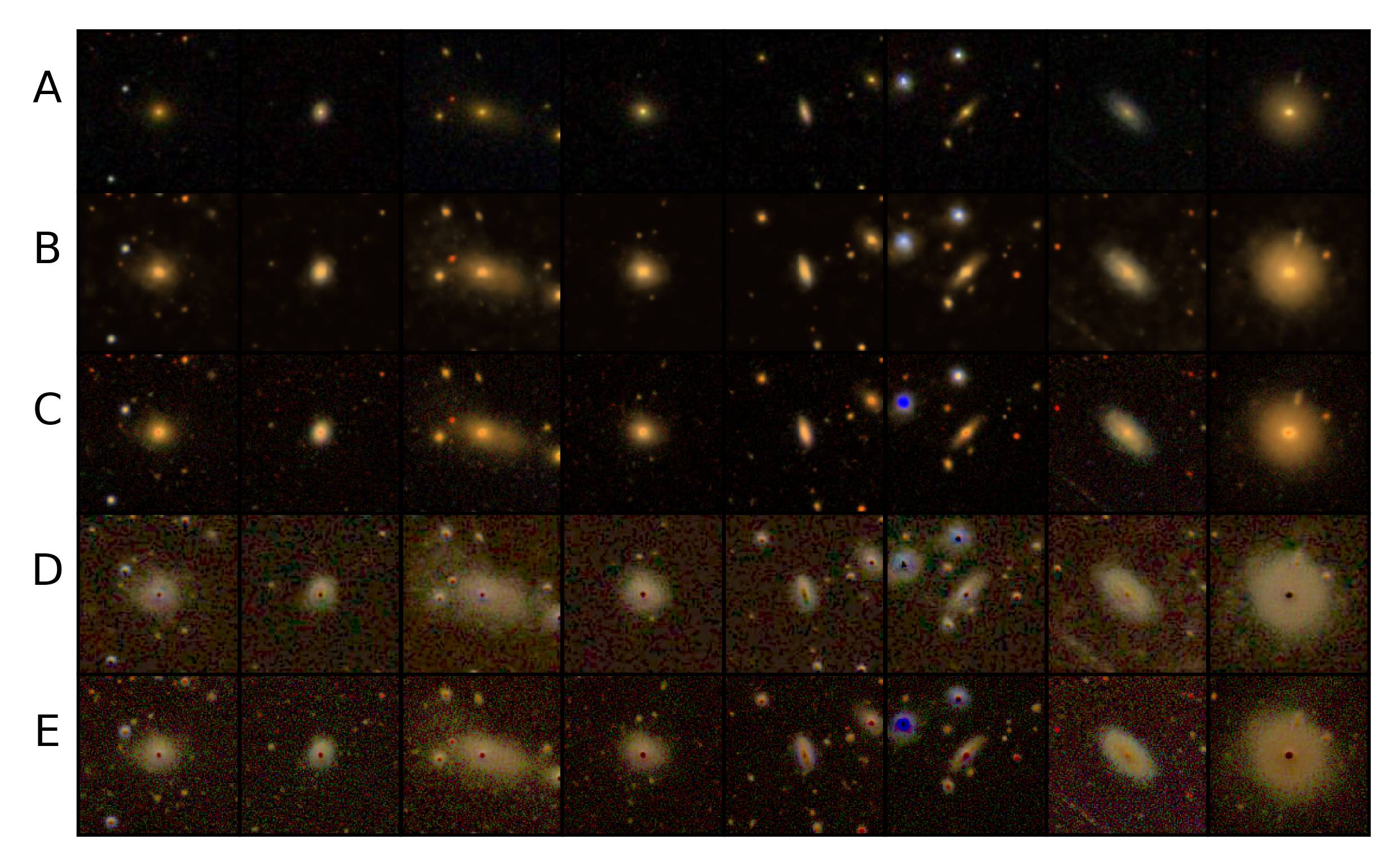}
    \end{minipage}
    \caption{Additional examples of source, target, and reconstructed images from the external dataset; the formatting is the same as in Figure \ref{fig:ExtCollage}.  Note that the images in rows D and E were enhanced for clarity.}
    \label{fig:ExtCollage5}
\end{figure*}

The rows in Figs. \ref{fig:Collage1} - \ref{fig:Collage6} represent the following quantities:

\begin{enumerate}[label=(\Alph*),labelwidth=*,align=left]
    \item SDSS representation
    \item DES representation
    \item CAE reconstruction
    \item CycleGAN reconstruction
    \item CAE residuals (CAE - DES)
    \item CycleGAN residuals (CycleGAN - DES)
    \item CAE gain (CAE - SDSS)
    \item CycleGAN gain (CycleGAN - SDSS),
\end{enumerate}

while in Fig. \ref{fig:ExtCollage5}, they represent the following quantities:

\begin{enumerate}[label=(\Alph*),labelwidth=*,align=left]
    \item SDSS representation
    \item CAE reconstruction
    \item CycleGAN reconstruction
    \item CAE gain (CAE - SDSS)
    \item CycleGAN gain (CycleGAN - SDSS),
\end{enumerate}

Note that to increase visibility, the residual and gain images were artificially enhanced with a power law transform (see Figs. \ref{fig:TestCollage} and \ref{fig:ExtCollage} for details).

\bsp	
\label{lastpage}
\end{document}